\documentclass[12pt]{article}
\pdfoutput=1 
\usepackage{amsmath,amssymb,amsthm}
\usepackage{bbm,latexsym}
\usepackage{graphicx}
\usepackage[numbers,sort&compress]{natbib}
\usepackage{authblk}
\usepackage{xcolor}
\usepackage{geometry}
\usepackage{hyperref}
\newcommand{\hc}{\mathrm{H.c.}}
\newcommand{\zed}{\mathbbm{Z}}

\newcommand{\mPl}{m_\mathrm{Pl}}

\begin{document}
\title{
\LARGE \bf Supersymmetric Majoron Inflation}
\setcounter{footnote}{2}
\author{Stephen F.\ King\thanks{E-mail: king@soton.ac.uk}\quad}
\author{\quad Patrick Otto Ludl\thanks{E-mail: P.Ludl@soton.ac.uk}}
\affil{{\small School of Physics and Astronomy, University of Southampton,\\Southampton, SO17 1BJ, United Kingdom}}

\date{\today}

\maketitle

\begin{abstract}
We propose supersymmetric Majoron inflation in which the Majoron field 
$\Phi$ responsible for generating right-handed neutrino masses may also be suitable for giving
low scale ``hilltop'' inflation,
with a discrete lepton number $\zed_{N}$ spontaneously broken at the end of inflation,
while avoiding the domain wall problem.
In the framework of non-minimal supergravity, we show that a successful
spectral index can result with small running together with small tensor modes.
We show that a range of heaviest right-handed neutrino masses can be generated,
$m_N\sim 10^1-10^{16}$ GeV, consistent with the constraints from reheating and domain walls.
\end{abstract}

\section{Introduction}

Following the discovery of neutrino oscillations~\cite{nobel}, we know that neutrinos have small masses with large mixing.
In the absence of a signal of neutrinoless double beta decay,
the nature of neutrino mass is unknown: they  
could be either Dirac or Majorana.\footnote{It is also possible that both Dirac and Majorana mass terms could appear
together in the Lagrangian.}
In the Standard Model (SM), where neutrinos are massless, the Lagrangian has an accidental global $U(1)_L$ symmetry,
corresponding to lepton number $L$ being conserved.\footnote{It also respects baryon number $B$,
corresponding to an accidental global $U(1)_B$ symmetry.}
If Majorana neutrino masses are introduced,
such terms would explicitly break lepton number $L$ by two units.
The origin of such light Majorana neutrino masses is unknown but 
it clearly is related to the question of the breaking of $L$.
In general $L$ may be broken explicitly or spontaneously~\cite{Gelmini:1980re}.
A commonly considered possibility for the origin of Majorana neutrino masses is
via the effective Weinberg operator~\cite{Weinberg:1979sa} which explicitly breaks $L$ by two units.
In the type-I seesaw model~\cite{seesaw}, the origin of the effective Weinberg operator is due to right-handed Majorana
neutrino masses, which would also break $L$ by two units. 
The origin of such right-handed neutrino masses is therefore also related to the breaking of lepton number.

In the framework of the type-I seesaw mechanism,
one possibility is to have right-handed neutrino mass terms appearing explicitly in the Lagrangian,
which would correspond to explicitly break lepton number.
This could be called ``soft breaking'' of $L$ since such mass terms have dimension 3, which is less than 4.
Another possibility is that $L$ is an exact global symmetry which is spontaneously broken by some 
scalar field $\Phi$, resulting in right-handed neutrino masses and a Goldstone boson called the Majoron~\cite{Chikashige:1980ui}.
The idea is to consider a scalar $\Phi$ with $L=2$ giving mass to the
right-handed neutrinos $N^c$ via the Yukawa interaction
\begin{equation}
 \mathcal{L}_{\Phi N N} = - \lambda \, \Phi \, \overline{N_R^c} N_R + \mathrm{H.c.}
\end{equation}
Since $\Phi$ acquiring a vev $\langle \Phi \rangle \neq 0$
breaks lepton number conservation,
one may call $\Phi$ the Majoron field, although more commonly this name is reserved for the associated 
Goldstone boson which arises when the global $U(1)_L$ symmetry is spontaneously broken.
This model is often referred to as the singlet Majoron model, and we shall refer to the complex scalar field $\Phi$
as the Majoron field. 

A supersymmetric (SUSY) singlet Majoron extension of the minimal supersymmetric
standard model (MSSM) has also been proposed~\cite{Giudice:1992jg} and
studied~\cite{Shiraishi:1993np,Espinosa:1995mh,Umemura:1993wc,Mohapatra:1993fe}.
The main focus of these studies has been on spontaneous $R$-parity violation.
However, we note that, in all singlet Majoron models, 
it seems unlikely that $L$ would be an exact global symmetry, since global
symmetries are not as protected as gauge symmetries. It seems more likely that in all such Majoron models,
supersymmetric or not, the lepton number 
$L$ would be an approximate global symmetry, broken by some higher order operators.
An exception would be models where the anomaly-free combination $U(1)_{B-L}$ is gauged,
but in this paper we shall not consider this possibility.

In this paper we consider a supersymmetric model in which global lepton number 
$L$ is explicity broken by higher order terms of the form $\Phi^n$ in the superpotential and potential.
Such terms explicitly break global $U(1)_L$ down to a discrete subgroup of lepton number
$\zed_N$, where $N=n$ if $n$ is odd and $N=2n$ if $n$ is even.
The resulting scalar potential (calculated in the framework of supergravity) 
leads to a vacuum expectation value of $\Phi$ which spontaneously breaks the
discrete lepton number $\zed_{N}$. 
We shall focus on the possibility that, before spontaneous symmetry breaking,
the scalar potential, including non-minimal K\"ahler corrections, is suitable for cosmological inflation
\cite{Guth:1980zm}.
This is interesting since it relates inflation to the mechanism responsible for the origin of neutrino masses.
In particular, the complex Majoron field $\Phi$ simultaneously provides 
inflation and, at the end of inflation at the global minimum, right-handed neutrino masses.
Majoron inflation has been studied recently in~\cite{Boucenna:2014uma}, without supersymmetry,
although higher order $\Phi^n$ terms were not considered and large tensor modes were shown to result
from chaotic type inflation. By contrast, here we shall use the higher order $\Phi^n$ terms to generate 
a model of new inflation (or ``hilltop inflation'')
very similar to the one presented in~\cite{Senoguz:2004ky} (see also~\cite{Boubekeur:2005zm}),
\textit{i.e.}\ the framework will be supergravity with non-minimal K\"ahler potential,
leading to small tensor modes.
It is also interesting to note that in our model the Majoron field $\Phi$ may also 
carry flavour symmetry quantum numbers and may be considered also to be 
a flavon, as for example in the model in \cite{King:2014iia}, where different right-handed neutrinos 
carry different $\zed_{N}$ charges. 
Flavon inflation has for example been considered in~\cite{Antusch:2008gw}.
There have also been other approaches which attempt to relate inflation to neutrino masses,
for example, our Lagrangian is similar to the one of~\cite{Antusch:2004hd}
where, however, the sneutrino component of $N^c$ is used as the inflaton,
while we use the Majoron scalar field $\Phi$.
Further approaches to relating inflation to neutrino masses have also been considered~\cite{Murayama:1992ua}.

The outline of this paper is as follows. In section~\ref{sec2}
we discuss the structure of our model and the resulting
scalar potential in the framework of supergravity with a non-minimal K\"ahler
potential. In section~\ref{sec3} we investigate how to avoid the domain wall problem
in our model and discuss the phenomenology of inflation.
There we also provide plots of the allowed parameter space
of our model for a selection of cutoff-scales $\Lambda$
and values $n$ in the higher-dimensional operator $\Phi^n/\Lambda^{n-2}$
in the superpotential.
Finally, we present our conclusions in section~\ref{concl}.

\section{The model with non-minimal K\"ahler potential}\label{sec2}

\begin{table}
	\centering
$$
\begin{array}{||c||ccccc||}
\hline \hline
&\hat{X}_0 & \hat{\Phi} & \hat{N}^c & \hat{H}_u & \hat{L} \\  \hline
\hline
\zed_{N} & 0 & 2  & -1 & 0 & 1 \\[2mm] \hline 
R&2&0&1&0 & 1 \\ 
\hline \hline
\end{array}
$$
\caption{\label{tab-S4}The lepton $\hat{L}$, Higgs $\hat{H}_u$, CP conjugated right-handed neutrino
$\hat{N}^c$, auxiliary $\hat{X}_0$ and inflaton (Majoron) $\hat{\Phi}$ superfields of the model,
and how they transform under the symmetries $\zed_{N}$ and $U(1)_R$.}
\label{model}
\end{table}

We consider a superpotential involving five superfields: An auxiliary field $\hat{X}_0$,
the inflaton (Majoron) field $\hat{\Phi}$, a CP conjugated right-handed neutrino field
$\hat{N}^c$, the Higgs field $\hat{H}_u$ and a lepton field $\hat{L}$.
The Higgs and Lepton doublets have usual Standard Model (SM) gauged electroweak quantum numbers, 
while the other superfields above are SM singlets.
We impose a discrete symmetry $\zed_{N}$ under which $\hat{X}_0$ and $\hat{H}_u$ 
carry zero charge, while $\hat\Phi$ has charge $2$,  $ \hat N^c$ has charge $-1$, and $\hat L$ has
charge $+1$, as shown in Table~\ref{model}.
It is clear from Table~\ref{model} that $\zed_{N}$ can be thought of as a discrete subgroup of lepton number $U(1)_L$.

In addition, we assume an $R$ symmetry, with charge assignments also shown in Table~\ref{model}.
The lowest order superpotential with $R=2$ allowed by the symmetries is then given by\footnote{If we include also the
Higgs field $\hat{H}_d$, there will be an additional term $\hat{X}_0 \hat{H}_u \hat{H}_d$ in the superpotential. As we will see later,
the K\"ahler potential of the model can always be chosen in such a way that all scalar fields apart from
the scalar component of $\hat{\Phi}$
vanish during inflation. Therefore, we do not need to take this additional coupling into account in this paper
and we will only consider couplings which involve the inflaton or the neutrino field. Note that the $R$ symmetry
forbids the $\mu$-term $\mu \hat{H}_u \hat{H}_d$. An effective $\mu$-term could be generated from the coupling
$\hat{X}_0 \hat{H}_u \hat{H}_d$ as described in~\cite{King:1997ia}.}
\begin{equation}
\mathcal{W} = \hat{X}_0 \left( \frac{\hat{\Phi}^n}{\Lambda^{n-2}} - M^2 \right) + \lambda \hat{\Phi} \hat{N}^c \hat{N}^c + y \hat{L} \hat{H}_u \hat{N}^c.
\label{W}
\end{equation}
For even $N$, the lowest power of $\hat{\Phi}$ which couples to $\hat{X}_0$ in the superpotential is given by 
$\hat{\Phi}^n$ with $n=N/2$, while for odd $N$ we have $n=N$. 
When the real scalar component of $\hat{\Phi}$ develops a vacuum expectation value (vev),
this will break the $\zed_{N}$ symmetry completely if $N$ is odd, or it will preserve a $\zed_{2}$
subgroup of $\zed_{N}$
if $N$ is even.

The mass parameters $\Lambda$, $M$, and 
the
dimensionless couplings $\lambda$ and $y$
can all be made real and positive
by rephasing of $\hat \Phi$, $\hat N^c$, $\hat X_0$ and $\hat{L}$, \textit{i.e.}\
without loss of generality we may assume
\begin{equation}
\Lambda>0,\quad M^2>0,\quad \lambda>0,\quad y>0.
\end{equation}

We start with the minimal K\"ahler potential
\begin{equation}
\mathcal{K} = |\hat{X}_0|^2 + |\hat{\Phi}|^2 + |\hat{N}^c|^2 + |\hat{H}_u|^2 + |\hat{L}|^2.
\end{equation}
The F-term scalar potential is then given by
\begin{equation}
V = e^{\mathcal{K}/\mPl^2} \left\{\sum_{i,j} (K^{-1})_{ij} D_{z_i}\mathcal{W} \left(D_{z_j}\mathcal{W}\right)^\ast
- 3\, \mPl^{-2} \, |\mathcal{W}|^2 \right\},
\label{V}
\end{equation}
\begin{equation}
D_{z_i}\mathcal{W} = \partial_{z_i} \mathcal{W} + \mPl^{-2} \mathcal{W}\, \partial_{z_i} \mathcal{K},\quad
K_{ij} = \frac{\partial^2 \mathcal{K}}{\partial z_i \partial z_j^\ast},
\end{equation}
where $z_i = X_0,\, \Phi,\, N^c,\, H_u,\, L$.\footnote{We denote the scalar component
of a superfield $\hat\phi$ by $\phi$. In particular, $z_i$ are the complex scalar components.} 
The reduced Planck mass $\mPl = 2.435 \times 10^{18}~\mathrm{GeV}$ is related to
Newton's constant via $\mPl^{-2} \equiv 8\pi G$.
The D-term contributions to the scalar potential are at least quartic
in the fields, so they do not contain any mass terms of fields.
Consequently, when we will discuss the masses of the fields
below, we can neglect the D-term contributions.
At the end of this discussion it will turn out that it is
possible to choose the K\"ahler potential in such a way
that all fields apart from $\Phi$ are zero during inflation.
As a consequence, since the D-term contributions are at least bilinear in
the gauge multiplet fields, they vanish during (and after) inflation.
Therefore, for our purposes, it is sufficient to study the
F-term scalar potential, \textit{i.e.}\ $V=V_F$ in this paper.

Let us now study the prerequisites for slow-roll inflation by computing the masses of the involved fields.
To do so,
we reformulate $V$ in terms of the ten real fields 
$\varphi = ( \mathrm{Re}\,X_0,\,\mathrm{Im}\,X_0,\,\mathrm{Re}\,\Phi,\,\mathrm{Im}\,\Phi,\,\mathrm{Re}\,N^c,\,\mathrm{Im}\,N^c,\, \mathrm{Re}\,H_u,\, \mathrm{Im}\,H_u,\, \mathrm{Re}\,L,\, \mathrm{Im}\,L )$.
The bilinear terms in the fields are then given by
\begin{equation}
\mathcal{L}_2 = \frac{1}{2} \sum_{i,j=1}^{10} (\mathcal{M}^2)_{ij} \varphi_i \varphi_j \equiv 
\frac{1}{2} \sum_{i,j=1}^{10} \frac{\partial^2 V(\varphi)}{\partial \varphi_i \partial \varphi_j} \Big\vert_{\varphi=0} \varphi_i \varphi_j.
\end{equation}
The squared-mass matrix $\mathcal{M}^2$ for our model is given by
\begin{equation}
\mathcal{M}^2 = \frac{2\,M^4}{\mPl^2} \mathrm{diag}(0,0,1,1,1,1,1,1,1,1),
\end{equation}
\textit{i.e.}\ with the minimal K\"ahler potential the real and imaginary parts of $X_0$ remain massless, while all other fields have a mass $\sqrt{2}\,M^2/\mPl$.
In new inflation models the fields are assumed to have small values (usually smaller than $\mPl$), such that
the potential is dominated by the constant term
\begin{equation}
V_0 = V(0) = M^4.
\end{equation}
In the slow-roll approximation (which we require to be
valid during inflation) the Hubble constant is determined by
\begin{equation}
H^2 \approx \frac{V}{3\,\mPl^2},
\end{equation}
which in our case becomes
\begin{equation}
H^2 \approx \frac{V_0}{3\,\mPl^2} = \frac{M^4}{3\,\mPl^2}.
\end{equation}
Fields with masses greater than the Hubble parameter
rapidly evolve to their minimum and are therefore not capable
of creating a long enough exponential expansion of the Universe.
In our simple model we have
\begin{equation}
\begin{split}
& m_{\mathrm{Re}\,\Phi}^2 = m_{\mathrm{Im}\,\Phi}^2 =
m_{\mathrm{Re}\,N^c}^2 =
m_{\mathrm{Im}\,N^c}^2 =\\
& m_{\mathrm{Re}\,H_u}^2 =
m_{\mathrm{Im}\,H_u}^2 =
m_{\mathrm{Re}\,L}^2 =
m_{\mathrm{Im}\,L}^2
= \frac{2\,M^4}{\mPl^2} > H^2,
\end{split}
\end{equation}
so none of the components of $\Phi$, $N^c$, $H_u$ and $L$ can be the inflaton.
Since, however, we are interested in a model based on
$\Phi$ as the inflaton, we have to modify the potential.

Motivated by our desire for $\Phi$ to be the inflaton, 
we consider 
a non-minimal K\"ahler potential~\cite{Senoguz:2004ky,Antusch:2004hd,Antusch:2008gw} for the five superfields $\hat{X}_0$,
$\hat{\Phi}$, $\hat{N}^c$, $\hat{H}_u$ and $\hat{L}$ which to order $\mPl^{-2}$ has the form,\footnote{The
$\zed_N$-symmetry of the model and the requirement of a real K\"ahler potential
would also allow adding terms of the form $\kappa_n \mPl^{2-n} (\hat{\Phi}^n + \hat{\Phi}^{\ast n})$.
This, however, will not be needed to tune the masses of the fields.}
\begin{equation}
\mathcal{K} = \sum_i |\hat{S}_i|^2 + \frac{1}{\mPl^2} \sum_{i<j} \kappa_{ij} |\hat{S}_i|^2 |\hat{S}_j|^2 + \frac{1}{\mPl^2} \sum_i \kappa_i |\hat{S}_i|^4,
\label{K}
\end{equation}
where $\hat{S_i} = \hat{X}_0, \hat{\Phi}, \hat{N}^c, \hat{H}_u, \hat{L}$.
Computing $V$ as before, $V_0=M^4$ is unchanged, but the squared masses become
\begin{equation}
m^2_{\mathrm{Re}\,X_0} = m^2_{\mathrm{Im}\,X_0} = -\frac{8 \kappa_{X_0} M^4}{\mPl^2}
\end{equation}
and
\begin{equation}
m^2_{\mathrm{Re}\,S} = m^2_{\mathrm{Im}\,S} = \frac{2M^4}{\mPl^2} (1-\kappa_{X_0 S}) \quad\quad (S=\Phi, N^c, H_u, L).
\end{equation}

We want the scalar field $\Phi$ (the Majoron) to be the inflaton.
This can lead to successful inflation since the potential involving $\Phi$ is particularly flat due to its
high power $n$ in the scalar potential.
In order to achieve this we need to ensure that, locally, all scalar fields apart from $\Phi$ have large enough positive mass squares so that they quickly roll to their zero field values, while $\Phi$ has a negative mass squared, and slowly rolls away from its zero field value. This is sometimes referred to as ``hilltop'' inflation.

In order to achieve this we suppose that, for all fields apart from $\Phi$,
\begin{equation}\label{kappa_condition1}
\kappa_{X_0} < -\frac{1}{24} \quad\text{and}\quad \kappa_{X_0 S} < \frac{5}{6} \quad\quad(S\neq \Phi).
\end{equation}
Then all fields except $\Phi$ will have masses larger than $H$ such that they rapidly evolve to their minima.
If the conditions in Eq.~(\ref{kappa_condition1}) are satisfied,
the squared-mass matrix at zero field value is positive definite for all fields except $\Phi$
and
\begin{equation}
\frac{\partial V}{\partial \varphi_i} \Big\vert_{\varphi_i=0} = 0 \quad\forall\,\varphi_i.
\end{equation}
Consequently, the minimum of $V$ the fields $S\neq \Phi$ will rapidly evolve to is $S=0$. Therefore, we can set $X_0=N^c=H_u=L=0$ during inflation.

Turning to the field $\Phi$ itself, we shall choose
\begin{equation}
\kappa_{X_0\Phi} > 1,
\end{equation}
so that the field $\Phi$ gets a negative mass term making $\Phi=0$ a local maximum of $V$. This allows
inflation with $\Phi$ slowly rolling to a local minimum at $\Phi \neq 0$.

Assuming $X_0 = N^c = H_u = L = 0 $ during inflation, the relevant potential
for the complex scalar field $\Phi$ becomes, using Eq.~(\ref{V}), with Eqs.~(\ref{W}) and~(\ref{K}),
\begin{equation}\label{fullpotential}
V(\Phi) =  \frac{\exp\left\{\frac{|\Phi|^2}{\mPl^2} \left( 1+ \kappa_\Phi \frac{|\Phi|^2}{\mPl^2} \right) \right\}}{1+\kappa_{X_0\Phi} \frac{|\Phi|^2}{\mPl^2}}
\left| \frac{\Phi^n}{\Lambda^{n-2}} - M^2 \right|^2.
\end{equation}
The Lagrangian for $\Phi$ is then given by
\begin{equation}
\begin{split}
\mathcal{L}(\Phi) & =
\frac{\partial^2\mathcal{K}}{\partial \Phi \partial \Phi^\ast} \Big\vert_{X_0=N^c=H_u=L=0}(\partial_\mu \Phi) (\partial^\mu \Phi)^\ast - V(\Phi)\\
& = \left(1+ 4\kappa_\Phi \frac{|\Phi|^2}{\mPl^2}\right) (\partial_\mu \Phi)(\partial^\mu \Phi)^\ast - V(\Phi).
\end{split}
\end{equation}
The assumption of a non-minimal K\"ahler potential thus leads to a non-canonically normalized
kinetic term in the Lagrangian.
However, since the effects of a non-minimal K\"ahler potential
on reheating are irrelevant, only
the spectral index $n_S$, the tensor-to-scalar ratio $r$
and the running of the spectral index may be subject to
relevant contributions from a non-minimal $\mathcal{K}$.
Since it turns out that our model can easily comply with
the observed values/bounds on these quantities for $\kappa_\Phi=0$,
we may avoid the complication of non-canonical normalization
by assuming $\kappa_\Phi$ to be small enough to neglect
its effect in the kinetic terms. Therefore, in the remainder of the paper
we will assume canonically normalized kinetic terms, \textit{i.e.}\
\begin{equation}
\mathcal{L}(\Phi) = (\partial_\mu \Phi) (\partial^\mu \Phi)^\ast - V(\Phi)
\end{equation}
with the potential $V$ of Eq.~(\ref{fullpotential}).

Since our inflation model is supersymmetric, $\Phi$ is necessarily complex,
and one might think that we would need to 
treat the model as a two-field inflation model, with the two real
fields being $\mathrm{Re}\,\Phi$ and $\mathrm{Im}\,\Phi$.
However, it is possible to show that, during the inflationary epoch,
the ratio $\mathrm{Im}\,\Phi / \mathrm{Re}\,\Phi$ is effectively frozen, with inflationary dynamics controlled
by the magnitude of the complex Majoron field $|\Phi|$.
This is explained in detail in appendices~\ref{appA} and~\ref{appB}.
The result is
a set of equations of motion for a single inflaton field $\phi \equiv \sqrt{2}\,|\Phi|$
which reads
\begin{subequations}
\begin{align}
& \ddot{\phi} + 3H \dot{\phi} + \frac{\partial V(\phi, \psi)}{\partial \phi} = 0,\\
& H^2 = \frac{1}{3\mPl^2}\left( V + \frac{1}{2} \dot{\phi}^2 \right),
\end{align}
\end{subequations}
where $\psi = \mathrm{Arg}\, \Phi = \mathrm{arctan}(\mathrm{Im}\,\Phi/\mathrm{Re}\,\Phi)$ and the derivative has to be evaluated
at the approximately constant value $\psi_0$ during inflation.

\section{Majoron inflation}~\label{sec3}

We now have the relevant prerequisites in order to study the domain wall problem
and the phenomenology of single field Majoron inflation in our model, which we will do
in this section, where the form of the potential $V(\phi, \psi)$ will be discussed.

\subsection{The domain wall problem}

In this subsection, we first discuss the conditions
for avoiding the domain wall problem~\cite{Zeldovich:1974uw} in our model.
The presently (and in the future)
observable Universe originates from a small
patch of the pre-inflationary Universe with
homogenous initial conditions $\Phi(t_0)$
in the whole patch. Since during
inflation there is an immense drop in temperature,
thermal fluctuations will not affect the time evolution
of $\Phi$. Consequently, $\Phi$ will approach the same
minimum everywhere in the Universe, therefore not
forming domains during inflation. The crucial question
is whether in the reheating phase of the Universe,
the temperature $T_R$ reaches a value higher
than the potential barrier between the $n$ equivalent
minima of the $\zed_n$-symmetric potential~(\ref{fullpotential}).
To answer this question, we have to compute the
height of the barrier and the reheating temperature,
which we will do in this section.
We will not discuss creation of domain walls due to quantum
fluctuations in this paper.

\subsubsection{The height of the barrier}

Reformulating the scalar potential~(\ref{fullpotential})
in terms of two real and positive fields $\phi$ and
$\psi$ defined as
\begin{equation}
\Phi = \frac{1}{\sqrt{2}}(\phi_R + i\phi_I) \equiv \frac{1}{\sqrt{2}} \phi \, e^{i\psi},
\end{equation}
we obtain
\begin{equation}\label{phipsipotential}
V(\phi,\psi) = f(\phi) g(\phi,\psi)
\end{equation}
with
\begin{equation}\label{functionf}
f(\phi) = \frac{\exp\left\{\frac{\phi^2}{4\mPl^2} \left( 2+ \kappa_\Phi \frac{\phi^2}{\mPl^2} \right) \right\}}{2^n \Lambda^{2n-4}\left(1+\kappa_{X_0\Phi} \frac{\phi^2}{2\mPl^2}\right)}, \quad
g(\phi,\psi) = (\phi^n-v_M^n)^2+2v_M^n\phi^n (1-\mathrm{cos}(n\psi)),
\end{equation}
where
\begin{equation}\label{mudef}
v_M \equiv \sqrt{2}(M^2 \Lambda^{n-2})^{1/n}
\end{equation}
is the vev of $\phi$.
The height of the barrier between two minima ($\mathrm{cos}(n\psi)=1$)
is thus given by
\begin{equation}
4f(\phi) \, v_M^n \phi^n.
\end{equation}
Since $\phi \ll \mPl$, we can expand $f(\phi)$
in $\mPl^{-1}$ giving
\begin{equation}
f(\phi) = \frac{1}{2^n\Lambda^{2n-4}} \left( 1 - \beta \frac{\phi^2}{\mPl^2}\right) + \mathcal{O}(\mPl^{-4}),
\end{equation}
where
\begin{equation}\label{defbeta}
\beta \equiv \frac{\kappa_{X_0\Phi}-1}{2} > 0.
\end{equation}
The height of the barrier is therefore given by
\begin{equation}
\Delta V(\phi) \approx \frac{v_M^n\phi^n}{2^{n-2}\Lambda^{2n-4}} = \frac{\sqrt{2}^{4-n}M^2}{\Lambda^{n-2}} \phi^n.
\end{equation}

\subsubsection{The reheat temperature}

We estimate the reheat temperature $T_R$ using the
prescription of~\cite{Kolb-Turner}.
For our purposes we only need to know the order
of magnitude of the reheat temperature and, therefore,
it is sufficient to treat also reheating as if our
model was a single-field inflation model.
For our computation we assume that the system
at the beginning of reheating already has evolved to
one of its minima with respect to $\psi=\mathrm{Arg}\,\Phi$,
in which case the potential~(\ref{phipsipotential}) becomes\footnote{Note that
in the minima with respect to $\psi$ one has $\Phi^n=\phi^n/\sqrt{2}^n$.}
\begin{equation}\label{Vapprox}
V(\phi) = f(\phi) (\phi^n-v_M^n)^2 \approx \frac{(\phi^n-v_M^n)^2}{2^n\Lambda^{2n-4}},
\end{equation}
with $v_M$ defined in Eq.~(\ref{mudef}).
The equation of motion for $\phi$ can then be recast as
\begin{subequations}
\begin{align}\label{single}
& \ddot{\phi} +3H\dot{\phi} + \frac{\partial V}{\partial \phi} = 0,\\
& H^2 = \frac{1}{3\mPl^2} \left( V + \frac{1}{2}\dot{\phi}^2 \right),
\end{align}
\end{subequations}
which is explained in detail in appendix~\ref{appB}.

Reheating happens through the decay of coherent oscillations
of the inflaton field (inflaton particles) to other particles
which subsequently thermalize. The equation of motion then
contains an additional friction term~\cite{Kolb-Turner} proportional
to the decay width $\Gamma_\phi$ of the inflaton:
\begin{equation}
 \ddot{\phi} + 3H\dot{\phi} + \Gamma_\phi \dot{\phi} + \frac{\partial V}{\partial\phi} = 0.
\end{equation}
Reheating therefore begins when the decay rate becomes comparable to the
expansion rate of the Universe, \textit{i.e.}\ for $\Gamma_\phi \gtrsim H$.
Inflaton decay proceeds via the Yukawa coupling
\begin{equation}
-\lambda \Phi \overline{N^c_R} N_R + \hc
\end{equation}
to the right-handed neutrinos. Assuming $m_N \ll m_\mathrm{Inf}$ the rate for the decay into (s)neutrinos
is given by\footnote{The textbook
formula for a 2-body decay has $8\pi$ in the denominator. Here we have identical
fermions/scalars in the final state, which yields an additional factor 1/2.
Another factor 1/2 comes from the fact that only the right-handed components of the
neutrinos couple to the inflaton. Finally, there is a factor 2, since also
the decay into sneutrinos is possible.}
\begin{equation}
\Gamma_\phi = \frac{\lambda^2 m_\mathrm{Inf}}{16\pi},
\end{equation}
where $m_\mathrm{Inf}$ is the inflaton mass.

Reheating starts at $\Gamma_\phi \sim H$ which implies
\begin{equation}\label{reheatingstart}
V + \frac{1}{2}\dot{\phi}^2 = 3 \mPl^2 H^2 \sim \frac{3 \mPl^2 \lambda^4 m_\mathrm{Inf}^2}{256\,\pi^2}.
\end{equation}
The left-hand side of this equation is just the energy density of the
scalar field. Assuming that once reheating starts it is almost completely
converted into thermal energy of the decay products, we find
\begin{equation}
\frac{g_\ast \pi^2}{30} T_R^4 \sim \frac{3 \mPl^2 \lambda^4 m_\mathrm{Inf}^2}{256\,\pi^2}.
\end{equation}
The reheat temperature is thus given by
\begin{equation}\label{Treh}
T_R^4 \sim \frac{45 \mPl^2 \lambda^4 m_\mathrm{Inf}^2}{128\,\pi^4 g_\ast},
\end{equation}
where $g_\ast$ is the number of (ultrarelativistic) degrees of freedom of the thermal bath
created by reheating.\footnote{Even if the mass of the right-handed (s)neutrinos is larger
than the reheat temperature a thermal bath can be created due to the decay
of the (s)neutrinos via the Yukawa coupling $y L H_u N^c$.}

\subsubsection{Creation of domain walls after inflation}

We finally want to compare the reheat temperature to the height of the
barrier between two minima during the reheating process. For this we need
the inflaton mass, which is obtained from the squared-mass matrix
\begin{equation}
(M_\phi^2)_{ij} = \frac{\partial^2 V}{\partial \phi_i \partial \phi_j}\Big\vert_{i,j=R,I;\, \phi=v_M} =
n^2 M^2 \left(\frac{M}{\Lambda}\right)^{2\left(1-\frac{2}{n}\right)} \delta_{ij}
\end{equation}
at the global minimum $\phi=v_M$ of the potential.
The inflaton mass is thus given by
\begin{equation}\label{minf}
m_\mathrm{Inf} = n M \left( \frac{M}{\Lambda} \right)^{1-\frac{2}{n}}.
\end{equation}
Note that in our model the masses of the scalar field $\phi$
and the pseudoscalar field $\psi$ are both equal to $m_\mathrm{Inf}$ at the global minimum of the potential.
Thus, there are two degenerate physical particles 
with common mass $m_{\phi}=m_{\psi}=m_\mathrm{Inf}$, which may be observable in future collider experiments,
if $m_\mathrm{Inf}$ is low enough.
The other information we need is the height $\Delta V$ of the potential
barrier at the beginning of reheating.
From Eq.~(\ref{reheatingstart}) we see that at the beginning of
reheating
\begin{equation}
V < \frac{3\mPl^2\lambda^4 m_\mathrm{Inf}^2}{256 \pi^2}.
\end{equation}
which, using approximation~(\ref{Vapprox}), yields
\begin{equation}
\phi^n > v_M^n - \frac{\sqrt{3\times 2^n} \lambda^2 \Lambda^{n-2} \mPl m_\mathrm{Inf}}{16\pi}
\end{equation}
\textit{i.e.}\
\begin{equation}
\Delta V > 4M^4 - \frac{n \sqrt{3} \lambda^2}{4\pi} \left( \frac{M}{\Lambda} \right)^{1-\frac{2}{n}} \mPl M^3.
\end{equation}
The creation of domain walls due to the thermal energy released by the reheating
process will be suppressed as long as
\begin{equation}
 \frac{\Delta V^{1/4}}{T_R} > 1,
\end{equation}
and thus the requirement for avoiding domain wall creation in
our model is
\begin{equation}\label{TRrequirement}
 \frac{\Delta V}{ T_R^4} = \frac{
4 \left( \frac{M}{\mPl} \right)^2 - \frac{n\sqrt{3}}{4\pi} \frac{M}{\mPl} a
}{
\frac{45n^2}{128} \frac{1}{\pi^4 g_\ast} a^2
} \gg 1,
\end{equation}
where we have introduced the abbreviation
\begin{equation}
a \equiv \lambda^2 \left( \frac{M}{\Lambda} \right)^{1-\frac{2}{n}}.
\end{equation}
In the limit\footnote{We assume the reheating temperature to be
higher than the top-quark mass, yielding the lower bound $g_\ast>(g_\ast)_\mathrm{SM}=106.75$.
Therefore, the right-hand side of inequality~(\ref{inequ1})
is larger than $\mathcal{O}(1000)/n$. Comparison to the finally obtained bound on
$a$---see Eq.~(\ref{arequirement})---thus justifies this approximation.}
\begin{equation}\label{inequ1}
\frac{n\sqrt{3}}{4\pi} \frac{M}{\mPl} a \gg \frac{45n^2}{128} \frac{1}{\pi^4 g_\ast} a^2
\Rightarrow \frac{\mPl}{M} a \ll \frac{32\pi^3 \sqrt{3}}{45 n} g_\ast,
\end{equation}
the condition~(\ref{TRrequirement}) simplifies to
\begin{equation}
4 \left( \frac{M}{\mPl} \right)^2 \gg \frac{n\sqrt{3}}{4\pi} \frac{M}{\mPl} a
\end{equation}
or
\begin{equation}\label{arequirement}
 \frac{\mPl}{M} a \ll \frac{16\pi}{n\sqrt{3}} \approx \frac{30}{n}.
\end{equation}
The condition for avoiding the domain wall problem in our model is
thus given by
\begin{equation}
\lambda^2 \left(\frac{M}{\Lambda}\right)^{1-\frac{2}{n}} \frac{\mPl}{M} \ll \frac{16\pi}{n\sqrt{3}}
\end{equation}
or
\begin{equation}
\lambda^2 \ll \frac{16 \pi}{n\sqrt{6}} \frac{v_M}{\mPl}.
\end{equation}
Therefore, the condition to avoid domain walls provides a bound on
the Yukawa coupling of the Majoron field to the right-handed neutrinos.
Interestingly, this bound depends only on the vev $v_M$ of the Majoron.

\subsection{Inflation phenomenology}

In order to compute the CMB observables,
we need to compute the slow-roll parameters which are given by
\begin{subequations}\label{slow-roll-param}
\begin{align}
& \epsilon = \frac{1}{2} \mPl^2 \left( \frac{V'}{V}\right)^2, \\
& \eta = \mPl^2 \frac{V''}{V}, \\
& \xi = \mPl^4 \frac{V' V'''}{V^2},\label{defxi}
\end{align}
\end{subequations}
where $'=\partial/\partial\phi$. The potential
is given by Eqs.~(\ref{phipsipotential})
and~(\ref{functionf}), where $\psi=\psi_0$,
the approximately constant value of the phase $\psi$
during inflation.
In the following we will show that
\begin{equation}
\epsilon \ll 1.
\end{equation}
In order to show this, we first show that during inflation
$\phi \ll v_M$. The slow-roll parameter $\eta$ for $\phi\ll v_M$
is given by
\begin{equation}
\eta \simeq -\frac{2n(n-1)\, \phi^{n-2}\, \mPl^2\, \mathrm{cos}(n\psi_0)}{v_M^n} -2\beta + \mathcal{O}(\mPl^{-2}),
\end{equation}
where $\beta$ is defined in Eq.~(\ref{defbeta}).
During inflation we must have $|\eta|<1$ which necessarily implies
\begin{equation}
|A\phi^{n-2}+2\beta|<1,
\end{equation}
where we have defined
\begin{equation}
A \equiv \frac{2n(n-1)\, \mPl^2\,\mathrm{cos}(n\psi_0)}{v_M^n}.
\end{equation}
If $\beta$ is not much larger than $\mathcal{O}(1)$ and $\mathrm{cos}(n\psi_0)$ is not accidentally close to zero, this leads to the upper
bound
\begin{equation}\label{phibound}
\frac{\phi}{v_M} \lesssim \left(\frac{v_M^2/\mPl^2}{2n(n-1)}\right)^{\frac{1}{n-2}}.
\end{equation}
This inequality necessarily holds also for $\phi_R$ and $\phi_I$
and thus
\begin{equation}
|\phi_R|^{n-2} \lesssim \frac{1}{2(n-1)} \frac{v_M^n}{n\,\mPl^2}.
\end{equation}
Comparing this equation to the condition~(\ref{singlefieldcondition2})
leads to the conclusion that the evolution of the ratio $\mathrm{Im}\,\Phi/\mathrm{Re}\,\Phi$
is always frozen in our model during inflation, and the assumption
of effective single-field inflation is justified.

Since $v_M$ as the vev of $\phi$ is the flavour symmetry breaking scale,
we assume that $v_M \lesssim M_\mathrm{GUT} \sim 10^{-3}\mPl$, in which case we find
\begin{equation}
n=3:\;\phi \lesssim 10^{-7}v_M,\quad
n=6:\;\phi \lesssim 10^{-2}v_M,\quad
n=9:\;\phi \lesssim 0.07 v_M.
\end{equation}
Therefore, for moderate $n$, \textit{e.g.} $n=6$, as anticipated $\phi \ll v_M$.

In the limit $\phi \ll v_M \ll \mPl$ the potential is given by\footnote{At the sample values
\begin{equation*}
v_M = 10^{-3}\,\mPl,\, \beta=0.05,\, \Lambda = \mPl,\, \mathrm{cos}(n\psi_0)=0.5, \, \kappa_\phi =1,
\end{equation*}
this approximation deviates from the exact form of $V$ by less than $10^{-4}\%$
in the range $\phi<0.1v_M$ for $n=3$. For higher $n$ the approximation becomes
even much better. We will therefore use the approximate form~(\ref{Vapprox2}) of $V$
for the remainder of the paper.}
\begin{equation}\label{Vapprox2}
V(\phi,\psi_0) \simeq M^4 \left( 1 - \beta \frac{\phi^2}{\mPl^2} -2\frac{\phi^n}{v_M^n}\mathrm{cos}(n\psi_0) \right).
\end{equation}
This shows the recognisable ``hilltop'' form of the potential, corrected by a Planck scale suppressed
term proportional to the parameter $\beta$.
Inserting the approximate expression for $V$ into the definitions of the slow-roll parameters,
and using the approximation $V\approx M^4$ during inflation,
one finds
\begin{equation}\label{epsilonfrac}
\epsilon = 2 \left( \beta \frac{\phi}{\mPl} + n \frac{\phi^{n-1}\mPl}{v_M^n} \mathrm{cos}(n\psi_0) \right)^2.
\end{equation}
For $\beta=1$, $\mathrm{cos}(n\psi_0)=1$, $v_M=10^{-3}\mPl$ and the upper bound~(\ref{phibound})
for $\phi$ one finds $\epsilon \sim 10^{-8}$ for $n=9$ and much smaller values for
smaller $n$, \textit{i.e.}\ $\epsilon$
effectively vanishes in our model.
Therefore, one prediction of our model is an unobservably small tensor to scalar
ratio $r = 16\,\epsilon$.

In order to compute the spectral index, we need to know
the field value $\phi_e$ at the end of inflation. Since $\epsilon$
is negligibly small, slow-roll inflation may end either at $|\eta|=1$
or $|\xi|=1$. However, it is possible to show that there is
a minimal value of $\beta$ for which $|\xi|$ can reach the value $1$,
which is given by
\begin{equation}
\beta = \frac{1}{\sqrt{(n-1)(n-2)}}.
\end{equation}
It will turn out that, in order to reproduce the correct
spectral index, $\beta$ must be (much) smaller than $0.1$, so
for $n$ smaller than 12, $|\xi|<1$ and the end of inflation is
characterized by $|\eta|=1$. Restricting ourselves to $\beta<1/2$
this gives
\begin{subequations}
\begin{align}
& \phi_e^{n-2} = \frac{1-2\beta}{A} \quad(\mathrm{cos}(n\psi_0)>0),\label{phie1}\\
& \phi_e^{n-2} = \frac{-1-2\beta}{A} \quad(\mathrm{cos}(n\psi_0)<0),\label{phie2}
\end{align}
\end{subequations}
\textit{i.e.}\
\begin{equation}
A\phi_e^{n-2} = \sigma-2\beta,
\end{equation}
where
\begin{equation}
\sigma \equiv \mathrm{sign}\,\mathrm{cos}(n\psi_0).
\end{equation}

\subsection{Number of e-folds and observables}
The number of e-folds between the epoch of horizon-exit of the scale $k_\ast = 0.002\,\mathrm{Mpc}$
at inflaton field value $\phi_\ast$
and the end of inflation at inflaton field value $\phi_e$ in the slow-roll approximation is given by
\begin{equation}
N_\ast = - \frac{1}{\mPl^2} \int_{\phi_\ast}^{\phi_e} \frac{V}{V'} d\phi.
\end{equation}
Using again the approximate potential~(\ref{Vapprox2}) and $V\approx M^4$
one obtains
\begin{equation}\label{efolds}
N_\ast = 
\frac{1}{2\beta (n-2)} \mathrm{ln} \frac{\phi^{n-2}}{2\beta (n-1) + A\phi^{n-2}} \Big\vert_{\phi_\ast}^{\phi_e}.
\end{equation}
This implies the consistency condition
\begin{equation}
2\beta(n-1) + A\phi^{n-2}>0
\end{equation}
which, according to
\begin{equation}
V'(\phi,\psi_0) \approx -\frac{M^4\phi}{(n-1)\mPl^2} \left( 2\beta (n-1) + A\phi^{n-2}\right),
\end{equation}
physically means
that $V'<0$ during the whole of inflation.
For positive $A$ (\textit{i.e.}\ $\mathrm{cos}(n\psi_0)>0$) this is fulfilled for every value of $\phi$.
For negative $A$ it implies the bound
\begin{equation}
\phi^{n-2}<\frac{2\beta(n-1)}{|A|}
\end{equation}
which, at the end of inflation, implies
\begin{equation}\label{condition_sign_A}
|A|\phi_e^{n-2} = 1+2\beta < 2\beta (n-1) \Rightarrow \beta > \frac{1}{2(n-2)}.
\end{equation}
Solving Eq.~(\ref{efolds}) for $\phi_\ast$ gives
\begin{equation}\label{equphiast}
A\phi_\ast^{n-2}
= 2(n-1)\beta \frac{1}{\frac{\sigma+(2n-4)\beta}{\sigma-2\beta}e^{2(n-2)\beta N_\ast}-1}.
\end{equation}
Note that for positive $A$, for every number of e-folds there is a solution as long as
$\phi_\ast$ is close enough to zero. For negative $A$, there is an upper bound for $N_\ast$
given by the condition that the $\phi_\ast$ of Eq.~(\ref{equphiast}) must be positive.
However, when in the following we will discuss the predictions for the spectral index,
it will turn out that, for moderate $n$, $\beta$ must be very close to zero, which means
that the condition~(\ref{condition_sign_A}) cannot be satisfied, and $\mathrm{cos}(n\psi_0)$
and thus also $A$
must be positive. We will therefore set $\sigma=+1$ for the remainder of the paper.

\begin{figure}
\begin{center}
\includegraphics[width=0.45\textwidth]{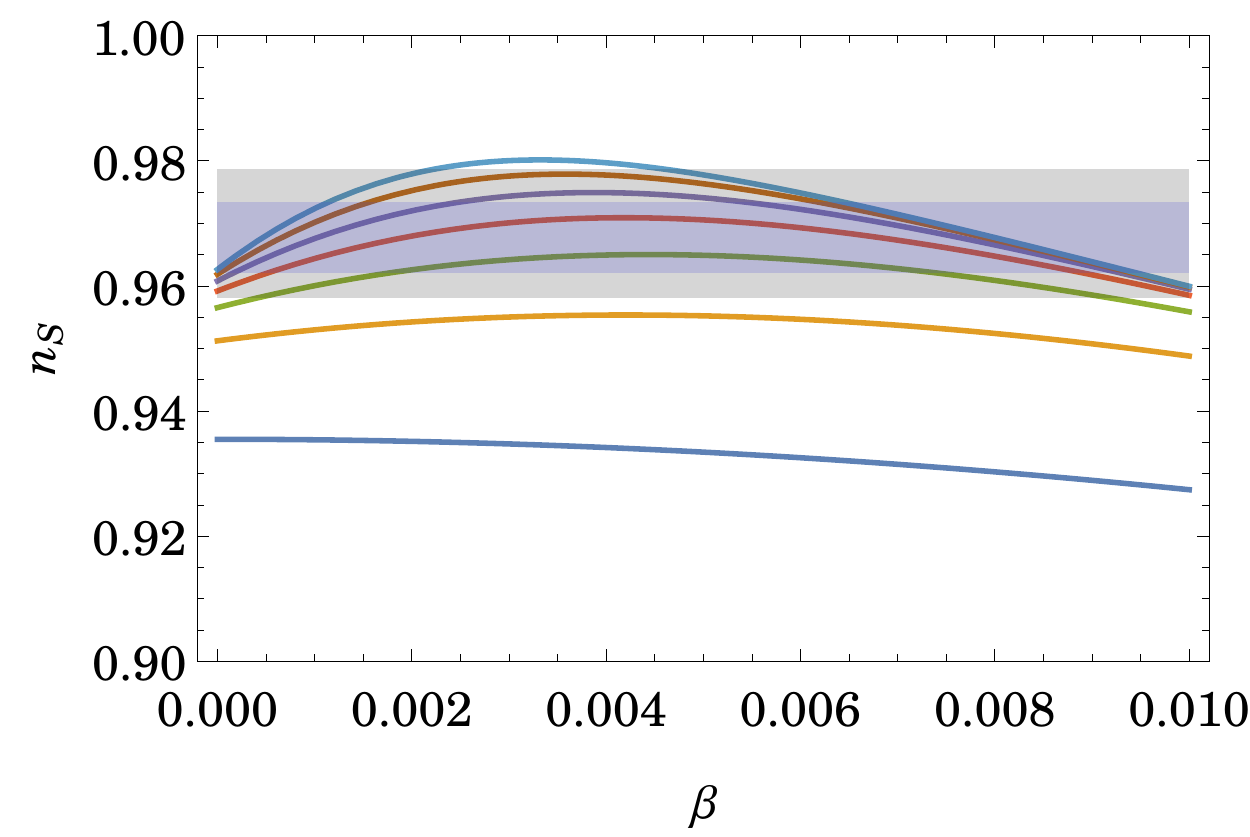}
\includegraphics[width=0.45\textwidth]{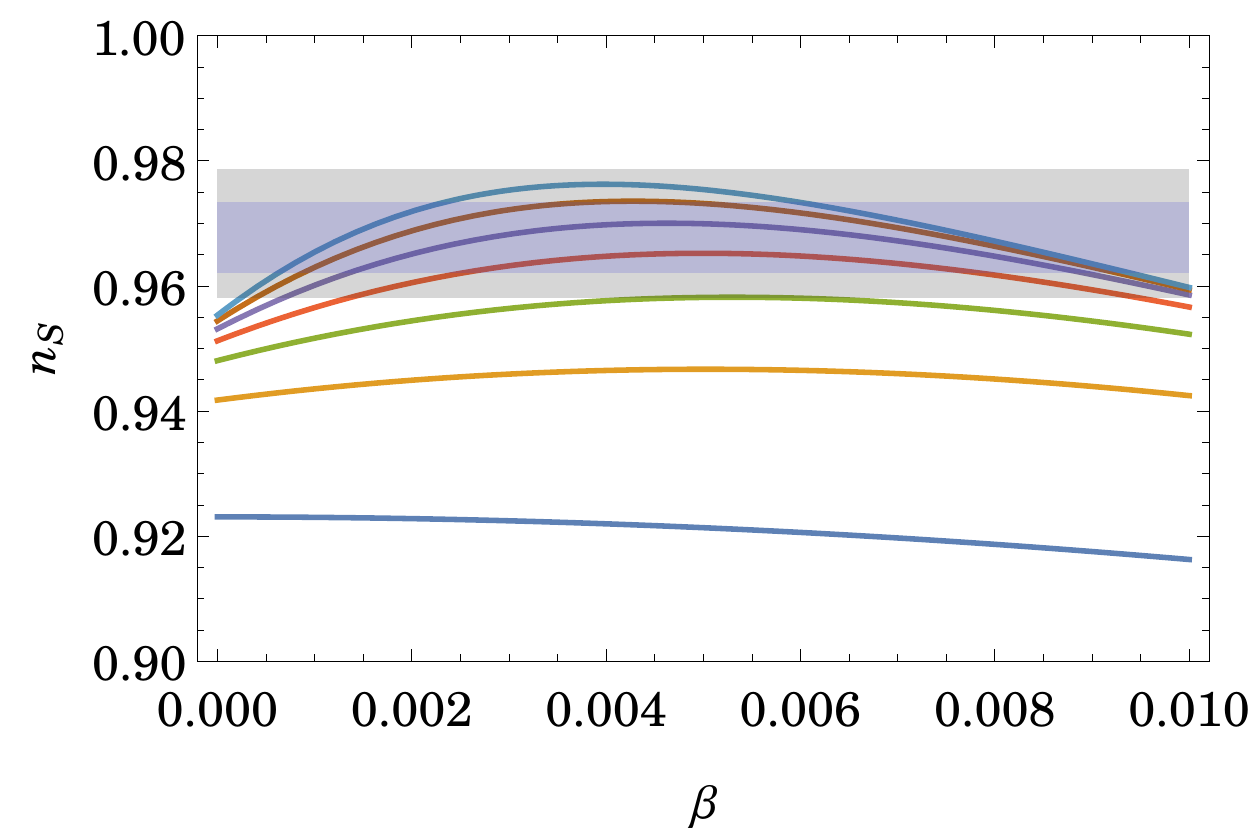}
\end{center}
\caption{The spectral index of our model as a function of $\beta$ for different values
of $n$ (lowest line: $n=3$, uppermost line: $n=9$).
The purple and gray bands are the 68\% and 95\% C.L. intervals for $n_S$
as determined by the Planck Collaboration in~\cite{Ade:2015xua} (Planck TT +lowP +BKP +lensing +ext).
Left: $N_\ast=60$, Right: $N_\ast=50$.}\label{fignS}
\end{figure}

The expressions
\begin{subequations}\label{observables}
\begin{align}
& n_S \approx 1 -6\epsilon_\ast +2\eta_\ast,\\
& r \approx 16 \epsilon_\ast,\\
& \frac{d\, n_S}{d\;\mathrm{ln}\,k}\Big\vert_{k_\ast} \approx -16 \epsilon_\ast \eta_\ast + 24 \epsilon_\ast^2 + 2\xi_\ast^2
\end{align}
\end{subequations}
for the CMB observables have to be evaluated at the field value $\phi_\ast$.
In this way, we obtain a relation between the number of e-folds
and $n_S$, $r$ and the running of the spectral index.
As discussed earlier, $\epsilon$ effectively vanishes in our model,
and the tensor to scalar ratio $r$ will be indistinguishable from zero.
Therefore, the interesting predictions will be the ones for $n_S$ and its running.
Since $\epsilon$ is negligibly small, we find
\begin{equation}
n_S = 1 + 2\eta_\ast = 1 -4\beta \left\{
1 + \frac{n-1}{\frac{1+(2n-4)\beta}{1-2\beta} e^{2(n-2)\beta N_\ast}-1}
\right\}.
\end{equation}
For $\beta\rightarrow 0$ and using $N_\ast \gg \mathcal{O}(1)$ one obtains the approximate relation
\begin{equation}
n_S \simeq 1- \frac{2(n-1)}{N_\ast(n-2)},
\end{equation}
which coincides with the result of~\cite{Senoguz:2004ky}.
Figure~\ref{fignS} shows the spectral index as a function of $\beta$
for different values of $N_\ast$.
The running of the spectral index, due to the smallness of $\epsilon$, is given by
\begin{equation}
\frac{d\, n_S}{d\;\mathrm{ln}\,k}\Big\vert_{k_\ast} \approx 2\xi_\ast^2.
\end{equation}
Using the approximate potential of Eq.~(\ref{Vapprox2})
and $V\approx M^4$ in the denominator of the definition~(\ref{defxi}),
and keeping only the lowest order term in $\phi/v_M$ and $\phi/\mPl$
one finds
\begin{equation}
\xi \approx 2\beta (n-2) A \phi^{n-2},
\end{equation}
\textit{i.e.}\
\begin{equation}
 \xi_\ast \approx 
\frac{4\beta^2 (n-1)(n-2) }{\frac{1+(2n-4)\beta}{1-2\beta}e^{2(n-2)\beta N_\ast}-1},
\end{equation}
where we have, as discussed for $\eta_\ast$, set $\sigma=+1$.
A numerical evaluation of $\xi_\ast$ shows that for $\beta\in[0,0.01]$, $n=3,\ldots,9$
and $N_\ast=60$ or $N_\ast=50$ one has
\begin{equation}
\frac{d\, n_S}{d\;\mathrm{ln}\,k}\Big\vert_{k_\ast} < 10^{-6}.
\end{equation}

\subsection{Examples: $n=6$ and $n=9$}

We will finally investigate two examples, $n=6$
(\textit{i.e.}\ a $\zed_{12}$-symmetry in the scalar and superpotential)
and $n=9$ (\textit{i.e.}\ a $\zed_9$-symmetry in the scalar and superpotential),
in the light of all the derived constraints.

\begin{figure}
\begin{center}
\includegraphics[width=0.48\textwidth]{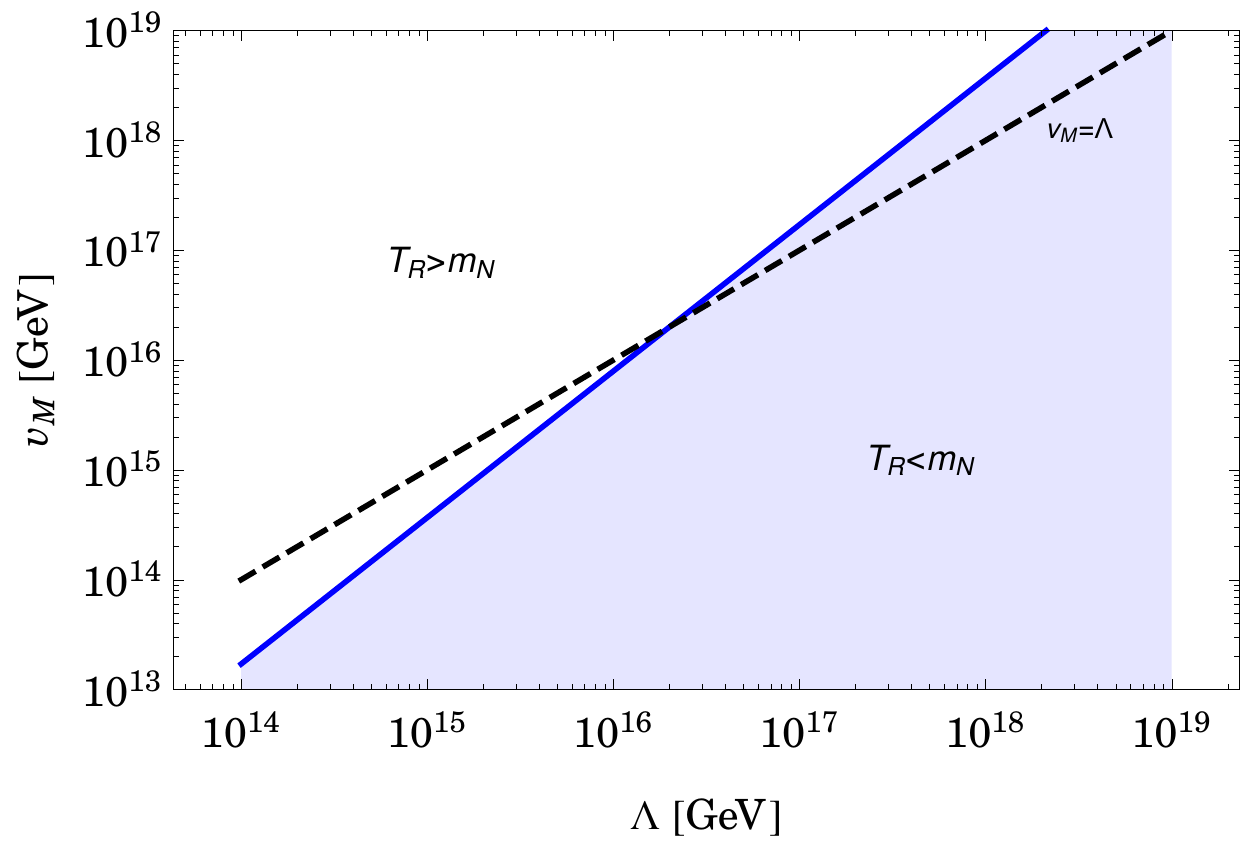}
\hspace{0.02\textwidth}
\includegraphics[width=0.48\textwidth]{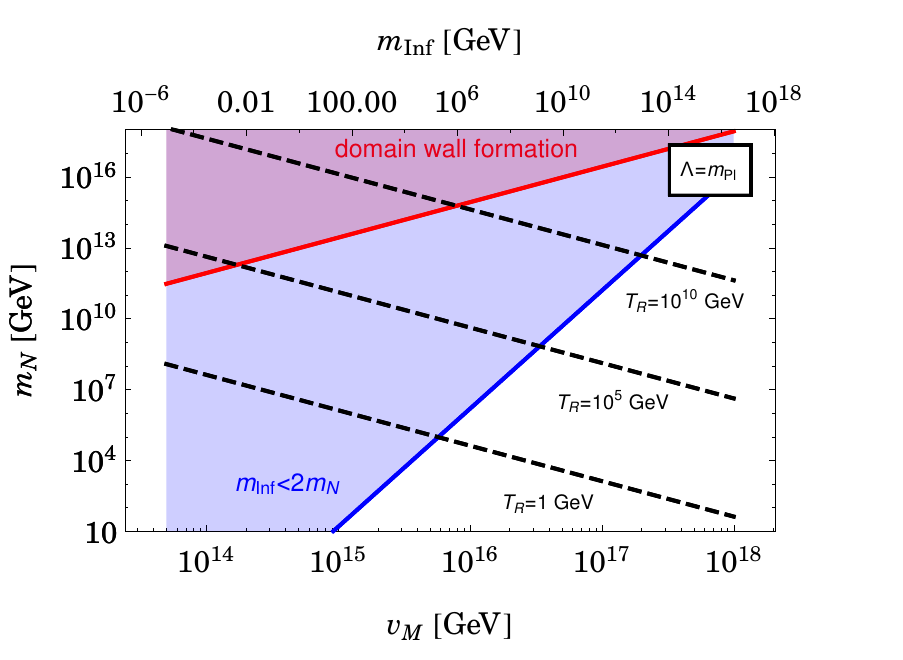}
\medskip
\\
\includegraphics[width=0.48\textwidth]{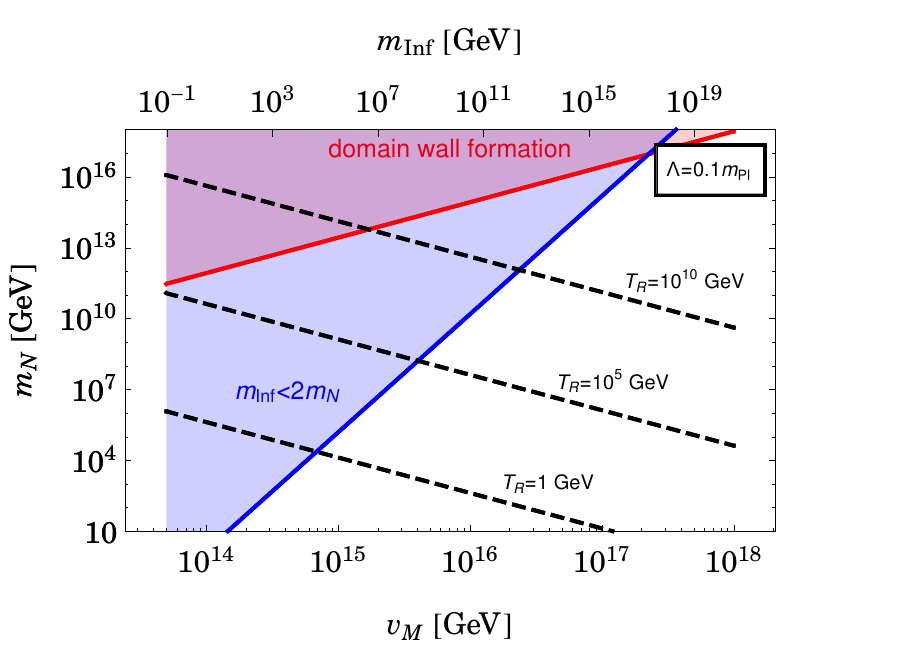}
\hspace{0.02\textwidth}
\includegraphics[width=0.48\textwidth]{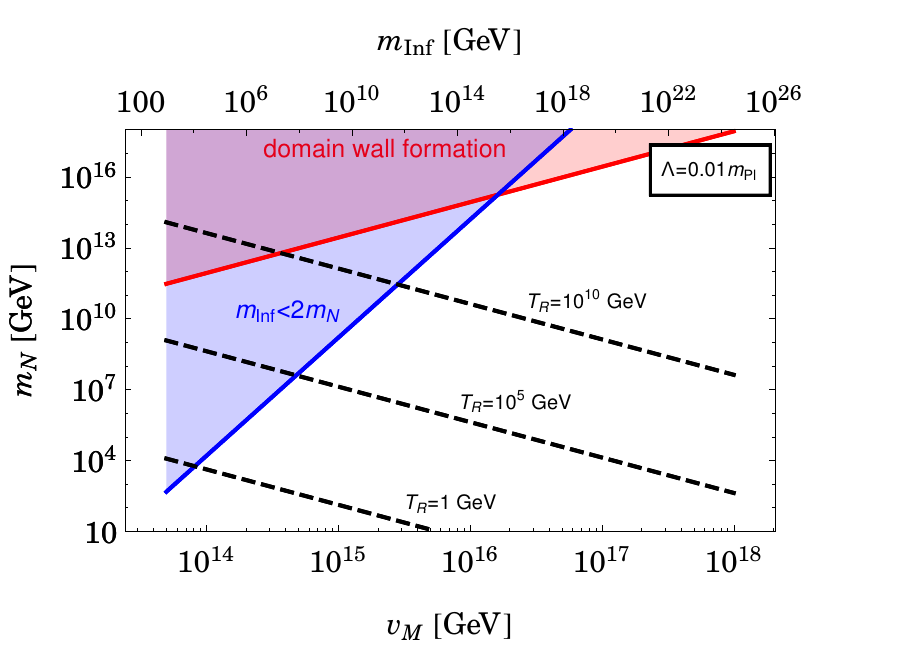}
\end{center}
\caption{The allowed regions (in white) of the parameter space for $n=6$.
\textit{Upper left:} This plot shows the values of $v_M$ as a function of $\Lambda$
such that $T_R>m_N$, \textit{i.e.}\ the right-handed neutrinos thermalize. The dashed black line
indicates $v_M=\Lambda$. Therefore, if we impose the flavour symmetry breaking scale to be lower than the
cutoff scale $\Lambda$, only the white area between the blue line and the dashed line
for $\Lambda\lesssim 10^{16}~\mathrm{GeV}$ is allowed. \textit{Upper right:}
The allowed parameter space (in white) for $n=6$ and $\Lambda=\mPl$.
The blue shaded area is excluded due to the kinematic condition $m_\mathrm{Inf}>2m_N$, and the
red shaded area is excluded due to domain wall formation. The three dashed lines
are the lines of constant reheat temperature $1~\mathrm{GeV}$,
$10^5~\mathrm{GeV}$ and
$10^{10}~\mathrm{GeV}$, respectively. \textit{Lower left and lower right:} The same
plots for $\Lambda=0.1\,\mPl$ and $\Lambda=0.01\,\mPl$,
respectively.}\label{fign6}
\end{figure}

We will start by asking the question whether
the right-handed neutrinos $N_R$ produced
via inflaton decay can be thermal. 
This will determine whether any subsequent leptogenesis is thermal or non-thermal.
In order for this
to be the case, we need
\begin{equation}
T_R > m_N,
\end{equation}
where
\begin{equation}
m_N = \frac{1}{\sqrt{2}} \lambda v_M
\end{equation}
is the right-handed neutrino mass.
From Eqs.~(\ref{Treh}) and~(\ref{minf})
one can derive the condition
\begin{equation}
\frac{\Lambda^{2n-4}}{v_M^{2n-6} \mPl^2} < \frac{45 n^2}{2^{n+4}\pi^4 g_\ast}
\end{equation}
for thermal right-handed neutrinos.
For $n=6$ this gives
\begin{equation}\label{eq1gast}
\frac{\Lambda^8}{v_M^6 \mPl^2} < \frac{405}{256 \pi^4 g_\ast} \approx 7\times 10^{-5},
\end{equation}
where we have used $g_\ast=240$ for the MSSM including three right-handed (s)neutrinos.
This is a very strong constraint, and it will usually not be fulfilled, unless the
flavour symmetry breaking scale $v_M$ and the cutoff scale $\Lambda$ are very close to
each other.
This means that typically at least the right-handed neutrino coupling to the inflaton
will be non-thermally produced. Consequently, $g_\ast<240$, but we assume that all MSSM
particles apart from one (or more) right-handed (s)neutrinos are produced thermally
and thus use the approximation $g_\ast \approx 240$ in the following.\footnote{The precise numerical value
of $g_\ast$ does not have any influence on the qualitative features of our model we
discuss in this section---see Eqs.~(\ref{eq1gast}) and~(\ref{eq2gast}). Therefore, all results obtained here are also valid
for reheat temperatures of the order of the top mass or smaller.}
The values of $\Lambda$ and $v_M$ for which, for $n=6$, the right-handed
neutrinos produced by inflaton decay thermalize are shown
in the upper left part of figure~\ref{fign6}.

Since we do, therefore, not impose a thermal $N_R$, the main condition for
reheating to be possible is the kinematical requirement
\begin{equation}
m_\mathrm{Inf} > 2 m_N,
\end{equation}
which is easily satisfied by an appropriate choice of $\lambda$.
The other constraint on the success of our model is the condition
$\Delta V \gg T_R^4$ for the successful avoidance of domain walls.

For these considerations only three physical parameters are relevant,
the cutoff scale $\Lambda$, the mass scale $M$ and the Yukawa coupling $\lambda$
of the superpotential. Fixing $\Lambda$ to a given value, the other two
quantities may be expressed in terms of
the flavour symmetry breaking scale (inflaton=Majoron vev) $v_M = \sqrt{2}(M^2\Lambda^{n-2})^{1/n}$
and the right-handed neutrino mass $m_N = \lambda v_M/\sqrt{2}$.
We will therefore show the allowed parameter regions of our model, for fixed $\Lambda$,
in a plot with the values of $v_M$ and $m_N$ shown on the axes.
Since the inflaton mass
\begin{equation}
m_\mathrm{Inf} = n M \left( \frac{M}{\Lambda}\right)^{1-\frac{2}{n}} = \frac{n}{\sqrt{2}} v_M \left(\frac{v_M}{\sqrt{2}\Lambda}\right)^{n-2} = [n=6] = \frac{3}{2\sqrt{2}} \frac{v_M^5}{\Lambda^4}
\end{equation}
is a function of $v_M$ and $\Lambda$ only, we may replace the $v_M$-axis by an
$m_\mathrm{Inf}$-axis.

The condition $m_\mathrm{Inf}>2 m_N$ leads to
\begin{equation}
 m_N < \frac{n}{2\sqrt{2}} v_M \left(\frac{v_M}{\sqrt{2}\Lambda}\right)^{n-2} = [n=6] = \frac{3}{4\sqrt{2}} \frac{v_M^5}{\Lambda^4},
\end{equation}
\textit{i.e.}\
\begin{equation}
\lambda < \frac{n}{2} \left( \frac{v_M}{\sqrt{2}\Lambda} \right)^{n-2} = [n=6] =
\frac{3}{4}\left( \frac{v_M}{\Lambda} \right)^4.
\end{equation}
In order to find the constraints on the parameter space with respect to
the condition $\Delta V \ll T_R^4$, we express the two quantities in
terms of $m_N$, $v_M$ and $\Lambda$:
\begin{subequations}
\begin{align}
& \Delta V = \Lambda^4 \left( \frac{v_M}{\sqrt{2}\Lambda} \right)^{2n} \left( 4 - \frac{\sqrt{6}n}{2\pi} \frac{\mPl m_N^2}{v_M^3} \right),\\
& T_R^4 = \Lambda^4 \left( \frac{v_M}{\sqrt{2}\Lambda} \right)^{2n} \frac{45 n^2}{16 \pi^4 g_\ast} \frac{\mPl^2 m_N^4}{v_M^6}.
\end{align}
\end{subequations}
The border to the region excluded by the domain wall problem is then given by
$\Delta V/T_R^4=1$ which can be expressed as
\begin{equation}\label{eq2gast}
m_N^2 = \frac{4\sqrt{6}\pi^2}{45n} \frac{v_M^3}{\mPl} \left( -g_\ast \pi + \sqrt{g_\ast^2 \pi^2 + 30 g_\ast}\right).
\end{equation}
For $n=6$ and $g_\ast=240$ this gives the bound
\begin{equation}
m_N < 1.31 \frac{v_M^{3/2}}{\mPl^{1/2}}.
\end{equation}
The upper right panel of figure~\ref{fign6} shows the allowed parameter space
for $n=6$ and $\Lambda=\mPl$. An evident feature is the
rather low reheat temperature for $v_M\lesssim M_\mathrm{GUT}$.
Therefore, if at least all SM particles
should be thermalized at the end of reheating, \textit{i.e.}\ $T_R\gtrsim 200~\mathrm{GeV}$,
we must have $v_M \gtrsim 10^{16}~\mathrm{GeV}$.
The lower half of figure~\ref{fign6} shows the same plots for $\Lambda=0.1\,\mPl$ and $\Lambda=0.01\,\mPl$.
For these scenarios the inflaton mass will be much larger and, consequently, much higher
right-handed neutrino masses $m_N$ are possible.

\begin{figure}
\begin{center}
\includegraphics[width=0.48\textwidth]{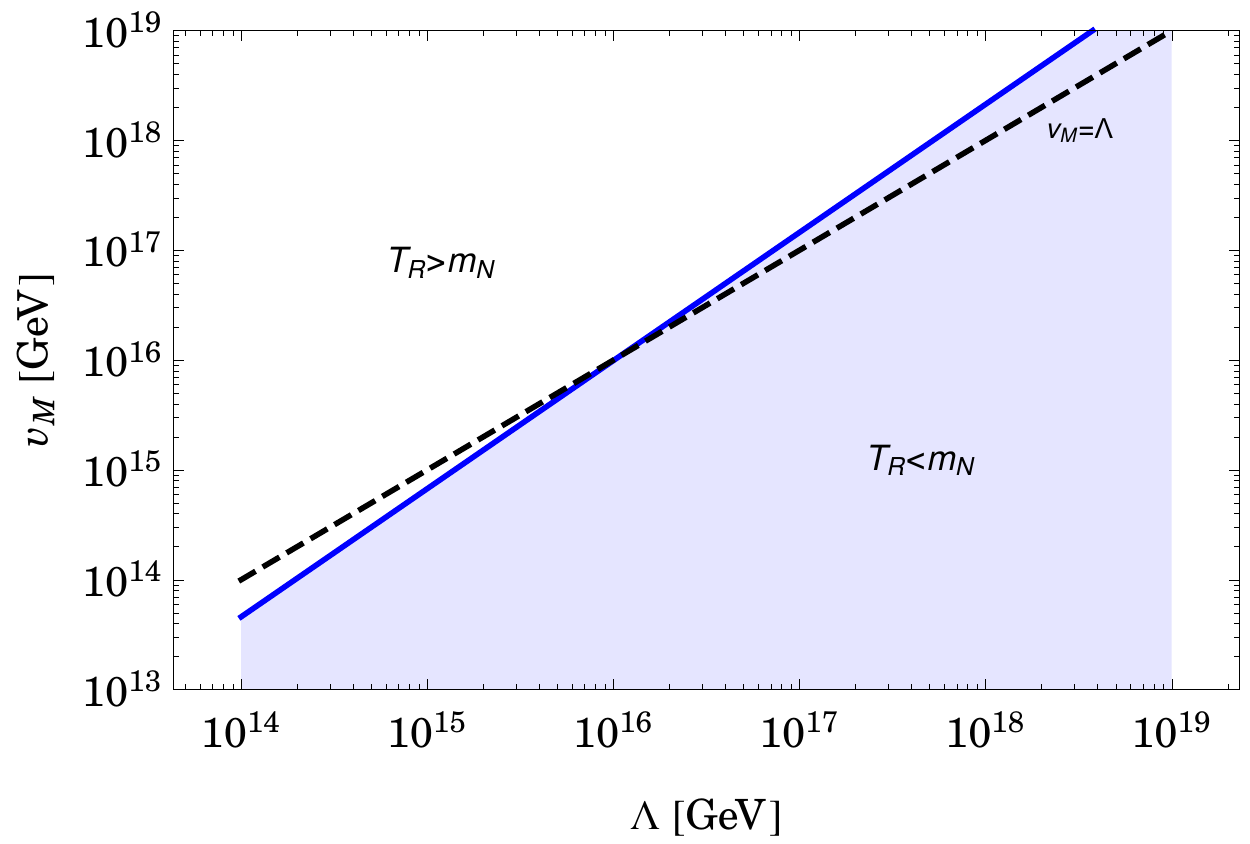}
\hspace{0.02\textwidth}
\includegraphics[width=0.48\textwidth]{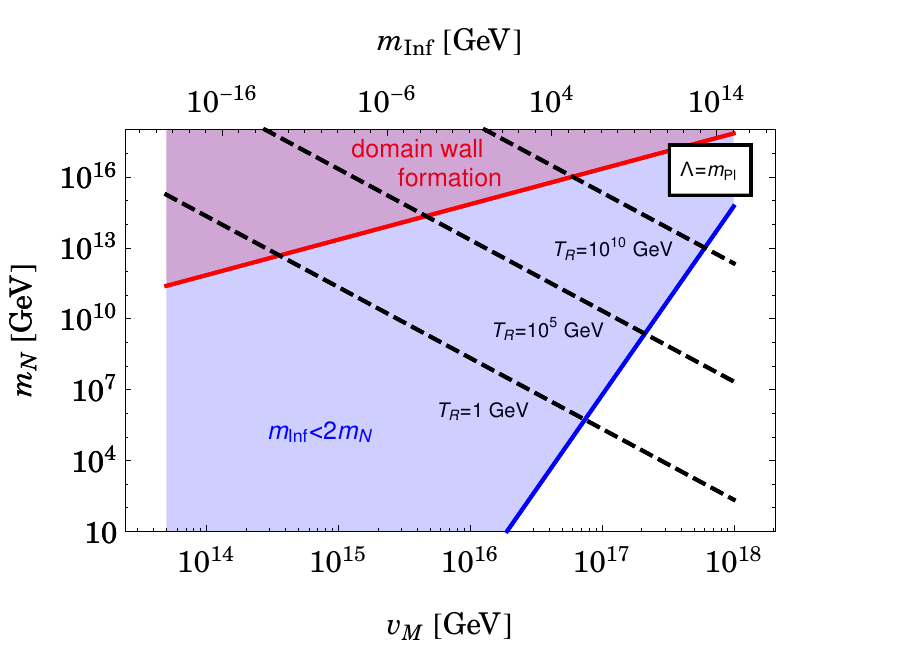}
\medskip
\\
\includegraphics[width=0.48\textwidth]{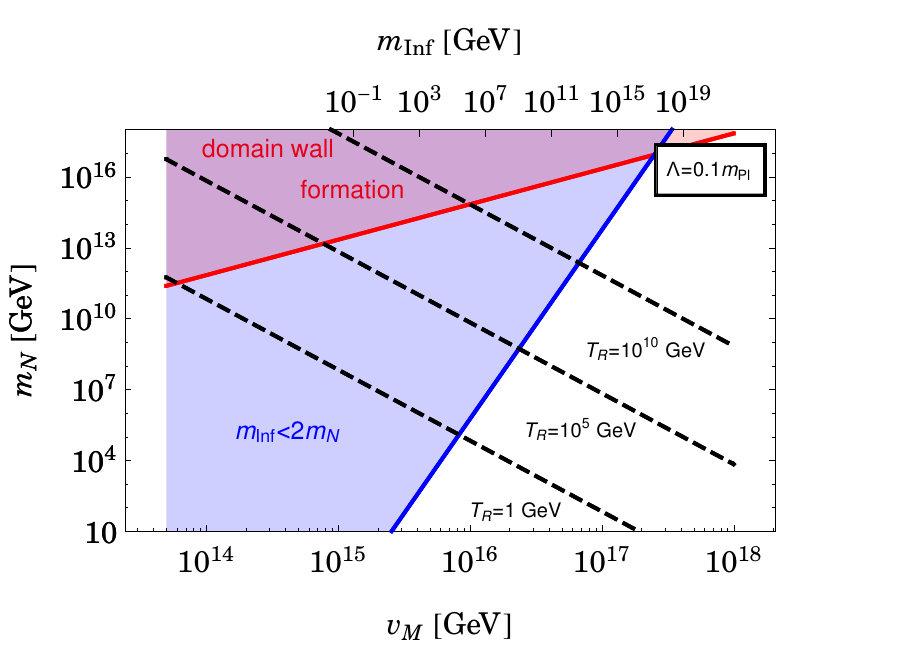}
\hspace{0.02\textwidth}
\includegraphics[width=0.48\textwidth]{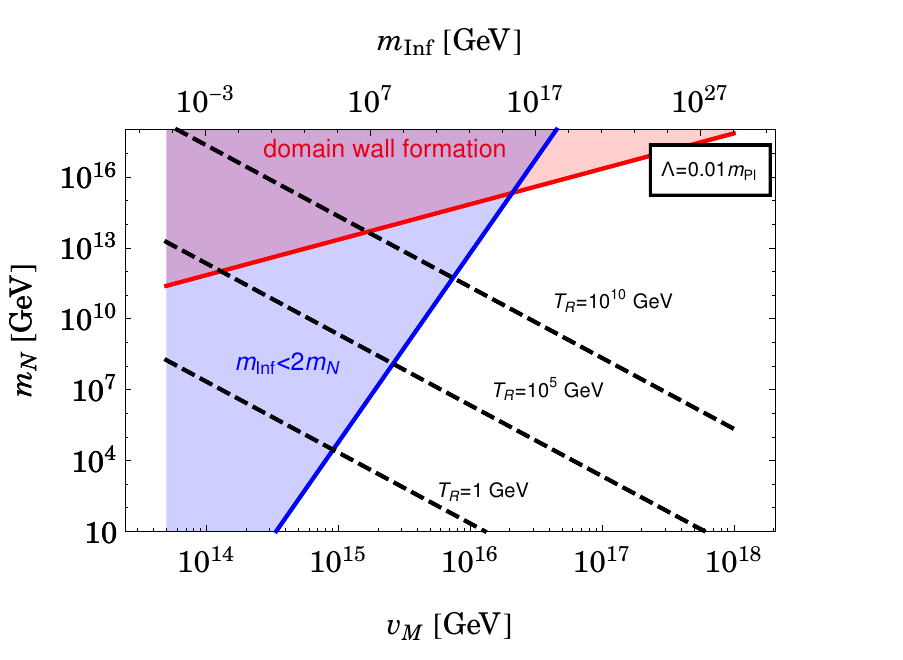}
\end{center}
\caption{The allowed regions (in white) of the parameter space for $n=9$.
\textit{Upper left:} This plot shows the values of $v_M$ as a function of $\Lambda$
such that $T_R>m_N$, \textit{i.e.}\ the right-handed neutrinos thermalize. The dashed black line
indicates $v_M=\Lambda$. \textit{Upper right:}
The allowed parameter space (in white) for $n=9$ and $\Lambda=\mPl$.
The blue shaded area is excluded due to the kinematic condition $m_\mathrm{Inf}>2m_N$, and the
red shaded area is excluded due to domain wall formation. The three dashed lines
are the lines of constant reheat temperature $1~\mathrm{GeV}$,
$10^5~\mathrm{GeV}$ and
$10^{10}~\mathrm{GeV}$, respectively. \textit{Lower left and lower right:} The same
plots for $\Lambda=0.1\,\mPl$ and $\Lambda=0.01\,\mPl$,
respectively.}\label{fign9}
\end{figure}

Finally, we also show the allowed parameter space
for $n=9$ in figure~\ref{fign9}. Qualitatively, figures~\ref{fign6}
and~\ref{fign9} look similar with, however, the parameter space
for $n=9$ being more constrained.

\newgeometry{left=1cm,right=1cm}
\restoregeometry

\newgeometry{left=1cm,right=1cm}
\restoregeometry

\subsection{Extension to more than one neutrino field}

Up to now we have considered only one neutrino field the inflaton
couples to. However, the generalization to more than one neutrino
is straight forward. Since reheating is the only
process related to inflation where the right-handed neutrinos
play a role, only the expression for $T_R$ will change.
Namely, if the neutrino-inflaton coupling is extended to
\begin{equation}
 \mathcal{L}_{\Phi N N} = - \sum_i \lambda_i \, \Phi \, \overline{N^c_{iR}} N_{iR} + \mathrm{H.c.}
\end{equation}
with neutrino mass eigenfields $N_i$, the decay width of the
inflaton will generalize to
\begin{equation}
\Gamma_\phi = \frac{m_\mathrm{Inf}}{16\pi} \left| \sum_i \lambda_i \right|^2.
\end{equation}
Consequently, the coupling $\lambda$ in the expression for the reheat temperature
has to be replaced by the effective coupling
\begin{equation}
\tilde{\lambda} \equiv \left| \sum_i \lambda_i \right|.
\end{equation}
Since in the framework outlined in this paper all right-handed neutrinos would get their masses via
the inflaton acquiring a vev, and since the right-handed neutrinos are usually
expected to have strongly hierarchical masses (due to the seesaw relation),
we will have
\begin{equation}
\tilde\lambda \approx \max_i |\lambda_i|,
\end{equation}
\textit{i.e.} the reheat temperature will, to a very good approximation,
be determined by the Yukawa-coupling of the heaviest right-handed neutrino.
The figures~\ref{fign6}
and~\ref{fign9} 
will therefore remain
unchanged, if $m_N$ on the y-axis is the mass of the heaviest right-handed neutrino.

\section{Conclusions}\label{concl}

In this paper we have considered a supersymmetric Majoron model in which the Majoron field 
$\Phi$ is responsible for both inflation and the generation of right-handed neutrino masses. 
In our model, global lepton number 
$U(1)_L$ is explicity broken to $\zed_N$ by high powers of the Majoron field $\Phi^n$ in the superpotential and potential.
Such terms explicitly break global $U(1)_L$ down to a discrete subgroup of lepton number
$\zed_N$ which is subsequently spontaneously broken by the vacuum expectation value 
of the Majoron, $v_M = \langle \Phi \rangle$.
We have focussed on the possibility that, before spontaneous symmetry breaking,
the scalar potential, including non-minimal K\"ahler corrections, is suitable for new or ``hilltop'' inflation.
This is interesting since it relates inflation to the mechanism responsible for the origin of neutrino masses.

Although $\Phi$ is a complex scalar, \textit{i.e.}\ with two real components, we have shown that, during inflation,
the ratio of imaginary and real part $\mathrm{Im}\,\Phi / \mathrm{Re}\,\Phi$ evolves slowly
compared to the expansion rate of the Universe,
so the model may be treated as a single
field inflation model.
We have discussed domain wall creation due to thermal fluctuations due to the reheating process,
and shown that, barring quantum fluctuations, after inflation the entire observable Universe would be expected to settle
into a single global minimum. We have computed the spectral index $n_S$
which depends on only one free parameter (from the K\"ahler potential), which can be tuned to be compatible
with the observed Planck value.
This agrees with the results of~\cite{Senoguz:2004ky} which uses a very similar superpotential
(the only effective difference with respect to the predictions for the CMB observables
being the restriction to even values of $n$).
We have shown that tensor modes are small as expected from ``hilltop'' inflation,
$r\approx 0$, and also the running of $n_S$ is negligible.

We investigated numerically two examples, $n=6$ corresponding to 
a discrete lepton number $\zed_{12}$
and $n=9$ corresponding to $\zed_9$.
For both examples, it turns out that the right-handed neutrino mass is larger than the reheat temperature $T_R$
for values of the cut-off scale $\Lambda$ above the GUT scale.
The inflaton may decay into pairs of right-handed neutrinos
providing its mass is large enough, $m_\mathrm{Inf}> 2m_N$, 
which may be achieved if $v_M = \langle \Phi \rangle$ is large enough (but not exceeding the cut-off $\Lambda$).
For both examples, we have shown that this may be achieved for a range of right-handed neutrino masses,
$m_N\sim 10^1-10^{16}$ GeV, where the lower bound on $m_N$ comes from requiring that $T_R \gtrsim 1$ GeV
and $v_M<\Lambda$,
and the upper bound on $m_N$ comes from requiring that domain walls are not formed at the end of inflation.\footnote{If
one would like to extend the present model to incorporate also leptogenesis, the reheat temperature
must be at least $\mathcal{O}(100)$ GeV in order to allow the sphaleron process. In this
case the lower bound for $m_N$ will become $m_N \gtrsim 10^3 - 10^4$ GeV instead of
$m_N \gtrsim 10$ GeV.}
We also considered extending the model to the case of three right-handed neutrinos and argued that,
if they have hierarchical masses, then the above
results will apply to the heaviest right-handed neutrino.

We note that in our model the mass of the physical scalar $\phi$
and the pseudoscalar $\psi$ components of the complex Majoron field $\Phi$
are both equal to $m_\mathrm{Inf}$ at the global minimum of the potential.
Thus, there are two degenerate physical particles 
with common mass $m_{\phi}=m_{\psi}=m_\mathrm{Inf}$, which in principle may be observable in future collider experiments,
if $m_\mathrm{Inf}$ is low enough. However, for $T_R \gtrsim 1$ GeV, we find $m_\mathrm{Inf}  \gtrsim 10^5$ GeV,
making them practically unobservable at present or planned future colliders.

Finally we mention that, since this is a new proposal,
there are inevitably several aspects of model building and cosmology which are beyond the scope
of this paper. In particular, we have not considered a complete flavour model from which such a Majoron inflation model
could emerge. In such a realistic model, the Majoron field here might carry additional flavour quantum numbers,
and there might be other fields in such models which could also play a role in cosmology.
We have also not considered the effects of reheating beyond our naive estimates, nor leptogenesis, which would most likely be non-thermal due to the fact that $T_R < m_N$ over most of the parameter space consistent with $v_M < \Lambda$.
These are all interesting aspects which are worth studying in the future.

In conclusion, we find that supersymmetric Majoron inflation is a promising and new idea which relates inflation
to neutrino masses via the type-I seesaw mechanism. Indeed we have shown that,
within the framework of non-minimal supergravity, the Majoron field 
$\Phi$ responsible for
generating right-handed neutrino masses may also be suitable for giving
low scale ``hilltop'' inflation, 
with a discrete lepton number $\zed_{N}$ spontaneously broken at the end of inflation,
while avoiding the domain wall problem.

\paragraph{Acknowledgements:} P.O.L.\ wants to thank Alexander Merle for helpful discussions.
The authors acknowledge support from the STFC grant
ST/L000296/1 and  the European Union Horizon 2020 research and innovation programme under
the Marie Sklodowska-Curie grant agreements
InvisiblesPlus RISE No.\ 690575 and Elusives ITN No.\ 674896.

\begin{appendix}

\section{Treatment of the model as a single-field inflation model}\label{appA}

Since our inflation model is supersymmetric, $\Phi$ is necessarily complex,
and we have to treat the model as a two-field inflation model, with the two real
fields being $\phi_R=\sqrt{2}\,\mathrm{Re}\,\Phi$ and $\phi_I=\sqrt{2}\,\mathrm{Im}\,\Phi$.\footnote{The
factors of $\sqrt{2}$ are introduced for convenience to obtain a Lagrangian
in terms of real fields which is canonically normalized: $\mathcal{L}=
\frac{1}{2} (\partial_\mu \phi_R)^2 + \frac{1}{2} (\partial_\mu \phi_I)^2
+ \ldots$~.}
The purpose of this appendix is to show that, during the inflationary epoch,
the ratio of these two fields is effectively frozen, with inflationary dynamics controlled
by the magnitude of the complex Majoron field $|\Phi|$.

We rewrite $\Phi$ in terms of $\phi_R$ and $\phi_I$, \textit{i.e.}\
\begin{equation}
\Phi \equiv \frac{1}{\sqrt{2}} \left( \phi_R + i\phi_I\right),
\end{equation}
which leads to
\begin{equation}
\mathcal{L} = \frac{1}{2}(\partial_\mu \phi_R)(\partial^\mu\phi_R) +
\frac{1}{2}(\partial_\mu \phi_I)(\partial^\mu\phi_I)
- V(\phi_R,\, \phi_I)
\end{equation}
From the energy-momentum tensor one then finds the energy density and pressure
of the scalar fields,
\begin{subequations}
\begin{align}
\rho = \frac{1}{2} \dot{\phi}_R^2 + \frac{1}{2} \dot{\phi}_I^2 + V(\phi_R,\, \phi_I),\label{scalarrho}\\
p = \frac{1}{2} \dot{\phi}_R^2 + \frac{1}{2} \dot{\phi}_I^2 - V(\phi_R,\, \phi_I),
\end{align}
\end{subequations}
and the equations of motion are given by
\begin{subequations}\label{EOM-twofield}
\begin{align}
& \ddot{\phi}_R + 3H \dot{\phi}_R + \frac{\partial V}{\partial \phi_R} = 0,\\
& \ddot{\phi}_I + 3H \dot{\phi}_I + \frac{\partial V}{\partial \phi_I} = 0,\\
& H^2 = \frac{1}{3\mPl^2} \left( V + \frac{1}{2}\dot{\phi}_R^2 + \frac{1}{2}\dot{\phi}_I^2 \right).
\end{align}
\end{subequations}
In the slow-roll approximation $\ddot{\phi}_{R,I} \ll 3H\dot{\phi}_{R,I}$, $\dot{\phi}_{R,I}^2 \ll V$
and one finds
\begin{subequations}
\begin{align}
& 3H \dot{\phi}_R \approx -\frac{\partial V}{\partial \phi_R},\label{EOMphiR}\\
& 3H \dot{\phi}_I \approx -\frac{\partial V}{\partial \phi_I},\label{EOMphiI}\\
& H^2 \approx \frac{V}{3\mPl^2}.
\end{align}
\end{subequations}
The value of the potential during the slow-roll phase is well approximated by
its value at vanishing field, \textit{i.e.} $V\approx V(0) \equiv V_0 = M^4$ and thus
\begin{equation}
H^2 \approx \frac{V_0}{3\mPl^2}.
\end{equation}
In the following we will show that during inflation the model can
effectively be treated as a single-field inflation model.
To do so, we study the time evolution of the ratio between the
imaginary and real part of $\Phi$,
\begin{equation}
\alpha \equiv \frac{\mathrm{Im}\,\Phi}{\mathrm{Re}\,\Phi} = \frac{\phi_I}{\phi_R},
\end{equation}
during inflation. The equation of motion for $\alpha$ is found
from Eqs.~(\ref{EOMphiR}) and~(\ref{EOMphiI}):
\begin{equation}
\frac{\dot{\alpha}}{\alpha} = \frac{\dot{\phi}_I}{\phi_I} - \frac{\dot{\phi}_R}{\phi_R} =
-\frac{1}{3H} \left( \frac{1}{\phi_I} \frac{\partial V}{\partial \phi_I} - \frac{1}{\phi_R} \frac{\partial V}{\partial \phi_R}\right).
\end{equation}
In new inflation models the field value is always much smaller than $\mPl$,
and we can, for the moment, safely set $\mPl\rightarrow\infty$. In this limit
we find
\begin{equation}\label{DGLalpha}
\frac{\dot{\alpha}}{\alpha} = \frac{n\,M^2}{3\sqrt{2}^n \Lambda^{n-2} H} \phi_R^{n-2} \frac{i}{\alpha} \left( (1+i\alpha)^n - (1-i\alpha)^n \right).
\end{equation}
The solution of this differential equation is given by
\begin{equation}\label{solutionalpha}
\int_{\alpha(t_0)}^\alpha \frac{d\alpha'}{i\left( (1+i\alpha')^n - (1-i\alpha')^n \right)}=
\frac{n\,M^2}{3\sqrt{2}^n} \int_{t_0}^t dt' \; \frac{1}{H(t')} \left(\frac{\phi_R(t')}{\Lambda}\right)^{n-2},
\end{equation}
where $t_0$ is the time when inflation starts and
$\alpha(t_0)$ the ratio of imaginary to real part of $\Phi$ in the small patch of the
pre-inflationary Universe which is inflated to the present (and in the future observable) Universe.
The integrand of the integral on the left-hand side of Eq.~(\ref{solutionalpha})
is the inverse of a polynomial in $\alpha$, and the integral therefore
has the form
\begin{equation}
\int_{\alpha(t_0)}^\alpha \frac{d\alpha'}{i\left( (1+i\alpha')^n - (1-i\alpha')^n \right)} = \mathrm{ln}\,\frac{F(\alpha)}{F(\alpha(t_0)))},
\end{equation}
where $F(\alpha)$ is a rational function of $\alpha$.
The solution may therefore be recast as
\begin{equation}
F(\alpha(t)) = F(\alpha(t_0)) \times \exp \left(
\frac{n\,M^2}{3\sqrt{2}^n} \int_{t_0}^t dt' \; \frac{1}{H(t')} \left(\frac{\phi_R(t')}{\Lambda}\right)^{n-2}
\right).
\end{equation}
Since the Hubble constant $H$ is, during inflation, also constant in time,
we find
\begin{equation}
\int_{t_0}^t dt' \; \frac{1}{H(t')} \left(\frac{\phi_R(t')}{\Lambda}\right)^{n-2} =
\frac{1}{\Lambda^{n-2} H(t_0)} \int_{t_0}^t dt' \; \phi_R(t')^{n-2}
\equiv \frac{t-t_0}{\Lambda^{n-2} H(t_0)} \overline{\phi_R}^{n-2},
\end{equation}
where we have defined an average field value $\overline{\phi_R}$ during inflation by
\begin{equation}
\overline{\phi_R}^{n-2} \equiv \frac{1}{t-t_0} \int_{t_0}^t dt' \; \phi_R(t')^{n-2}.
\end{equation}
In this way we find the following expression for the time evolution of $\alpha$
during inflation:
\begin{equation}
F(\alpha(t)) = F(\alpha(t_0)) \times \exp \left( \frac{n\,M^2}{3\sqrt{2}^n \Lambda^{n-2} H} \overline{\phi_R}^{n-2} (t-t_0) \right).
\end{equation}
Therefore, if
\begin{equation}\label{singlefieldcondition}
\frac{n\,M^2 |\overline{\phi_R}|^{n-2}}{3\sqrt{2}^n\Lambda^{n-2} H} < H
\end{equation}
during inflation,
the evolution of $\alpha$ is slow compared to the expansion
time scale and we can treat our model as a single field inflation model
by setting $\alpha(t)\approx \alpha(t_0)$ and making the replacement
\begin{equation}
\phi_I \approx \alpha(t_0) \phi_R.
\end{equation}
Using $H^2\approx M^4/3\mPl^2$, the condition~(\ref{singlefieldcondition})
becomes
\begin{equation}\label{singlefieldcondition2}
|\overline{\phi_R}|^{n-2} < \frac{v_M^n}{n\,\mPl^2},
\end{equation}
where
\begin{equation}
v_M = \sqrt{2} (M^2\Lambda^{n-2})^{1/n}.
\end{equation}
However, also the converse situation
\begin{equation}
|\overline{\phi_R}|^{n-2} > \frac{v_M^n}{n\,\mPl^2},
\end{equation}
leads to an effective single-field inflation model.
Namely, in this case, during the first few e-folds of
inflation $\alpha$ rapidly approaches one of its asymptotic values
(depending on the evolution of $\overline{\phi_R}(t)$)
\begin{equation}
\alpha_\infty = \tan(2\pi k/n) \quad (k\in\{0,\ldots,n-1\}).
\end{equation}
This corresponds to the $n$ equivalent minima of the $\zed_n$-symmetric potential, which in the complex
plane lie in the directions of the real axis and the directions defined by the non-trivial $n$th roots of
unity $\exp(2\pi i k/n)$.
Hence, in this scenario, the field can be assumed to be in one of its minima (with respect to $\alpha$)
during inflation which means
\begin{equation}
\phi_I \approx \alpha_\infty \phi_R,
\end{equation}
again allowing to treat the model as a model of single-field inflation.

The equations of motion for the effective single-field inflation
model are derived in appendix~\ref{appB}. The result is
a set of equations of motion for $\phi = \sqrt{\phi_R^2 + \phi_I^2}$
which reads
\begin{subequations}
\begin{align}
& \ddot{\phi} + 3H \dot{\phi} + \frac{\partial V(\phi, \psi)}{\partial \phi} = 0,\\
& H^2 = \frac{1}{3\mPl^2}\left( V + \frac{1}{2} \dot{\phi}^2 \right),
\end{align}
\end{subequations}
where $\psi = \mathrm{arctan}\,\alpha$ and the derivative has to be evaluated
at the approximately constant value $\psi_0$ during inflation.

\section{Derivation of the equation of motion in the effective
single-field inflation framework}\label{appB}

The equations of motion for the two fields $\phi_R$
and $\phi_I$ defined as
\begin{equation}
\Phi \equiv \frac{1}{\sqrt{2}} \left( \phi_R + i \phi_I \right) \equiv \frac{1}{\sqrt{2}} \phi e^{i\psi}
\end{equation}
are
\begin{subequations}\label{EQsys}
\begin{align}
& \ddot{\phi}_R + 3H \dot{\phi}_R + \frac{\partial V}{\partial \phi_R} = 0,\label{EOMa}\\
& \ddot{\phi}_I + 3H \dot{\phi}_I + \frac{\partial V}{\partial \phi_I} = 0,\label{EOMb}\\
& H^2 = \frac{1}{3\mPl^2} \left( V + \frac{1}{2}\dot{\phi}_R^2 + \frac{1}{2}\dot{\phi}_I^2 \right),\label{EOMc}
\end{align}
\end{subequations}
\textit{i.e.}\ Eqs.~(\ref{EOM-twofield}).
The equations of motion for the effective single-field inflation
model can be derived by rewriting Eqs.~(\ref{EQsys})
in terms of the polar coordinates $\phi$ and $\psi$:
\begin{equation}
\begin{split}
& \phi_R = \phi \cos\psi,\quad \phi_I = \phi \sin \psi,\\
& \frac{\partial}{\partial \phi_R} =
\cos\psi \frac{\partial}{\partial \phi} - \sin \psi \frac{1}{\phi}\frac{\partial}{\partial \psi}, \quad
\frac{\partial}{\partial \phi_I} =
\sin\psi \frac{\partial}{\partial \phi} + \cos \psi \frac{1}{\phi}\frac{\partial}{\partial \psi}.
\end{split}
\end{equation}
Adding $\cos\psi$ times Eq.~(\ref{EOMa}) to $\sin\psi$ times Eq.~(\ref{EOMb}) then
reveals
\begin{equation}
\ddot{\phi} + 3H\dot{\phi} + \frac{\partial V(\phi,\psi)}{\partial \phi} - \phi \dot{\psi}^2 = 0,
\end{equation}
and similarly one obtains
\begin{equation}
H^2 = \frac{1}{3\mPl^2} \left( V + \frac{1}{2} \dot{\phi}^2 + \frac{1}{2} \phi^2 \dot{\psi}^2 \right).
\end{equation}

In the case where $\alpha = \phi_I/\phi_R$ evolves slowly
compared to the expansion rate, or is already close to its
asymptotic value---see the discussion in appendix~\ref{appA}---we have
$\dot{\alpha}\approx 0$ and hence $\dot{\psi}\approx 0$ and the
equations of motion reduce to the following set of equations of motion
for a single real scalar field $\phi = \sqrt{2} |\Phi|$:
\begin{subequations}
\begin{align}
& \ddot{\phi} +3H\dot{\phi} + \frac{\partial V}{\partial \phi} = 0,\\
& H^2 = \frac{1}{3\mPl^2} \left( V + \frac{1}{2}\dot{\phi}^2 \right).
\end{align}
\end{subequations}
This is exactly the form expected for a single-field inflation model.
The derivative $\partial V/\partial \phi$ is to be taken at $\psi = \psi_0$,
where $\psi_0$ is the (approximately constant) value of $\psi$ during inflation.
\end{appendix}


\begin{thebibliography}{99}

\bibitem{nobel}
Special Issue on
``Neutrino Oscillations: Celebrating the Nobel Prize in Physics 2015''
Edited by Tommy Ohlsson,
Nucl.\ Phys.\ B {\bf 908} (2016) Pages 1-466 (July 2016),\\
\url{http://www.sciencedirect.com/science/journal/05503213/908/supp/C}.

\bibitem{Gelmini:1980re}
  G.~B.~Gelmini and M.~Roncadelli,
  Phys.\ Lett.\  {\bf 99B} (1981) 411.
  
\bibitem{Weinberg:1979sa}
  S.~Weinberg,
  Phys.\ Rev.\ Lett.\  {\bf 43} (1979) 1566.
  
   \bibitem{seesaw}
  P.~Minkowski,
  Phys.\ Lett.\ B {\bf 67} (1977) 421;
  T.~Yanagida in {\it Proc. of the Workshop on Unified Theory and
Baryon Number of the Universe}, KEK, Japan (1979);
%
M.~Gell-Mann, P.~Ramond and R.~Slansky in Sanibel Talk,
CALT-68-709, Feb 1979, and in {\it Supergravity}, North Holland,
Amsterdam (1979);
S.~L.~Glashow, Cargese Lectures (1979);
%
R.~N.~Mohapatra and G.~Senjanovic,
Phys.\ Rev.\ Lett.\  {\bf 44} (1980) 912.

  
  
\bibitem{Chikashige:1980ui}
  Y.~Chikashige, R.~N.~Mohapatra and R.~D.~Peccei,
  Phys.\ Lett.\  {\bf 98B} (1981) 265.

\bibitem{Giudice:1992jg}
  G.~F.~Giudice, A.~Masiero, M.~Pietroni and A.~Riotto,
  Nucl.\ Phys.\ B {\bf 396} (1993) 243
  [hep-ph/9209296].
  
\bibitem{Shiraishi:1993np}
  M.~Shiraishi, I.~Umemura and K.~Yamamoto,
  Phys.\ Lett.\ B {\bf 313} (1993) 89.
  
  

\bibitem{Espinosa:1995mh}
  J.~R.~Espinosa,
  Phys.\ Lett.\ B {\bf 353} (1995) 243
  [hep-ph/9503255].

\bibitem{Umemura:1993wc}
  I.~Umemura and K.~Yamamoto,
  Nucl.\ Phys.\ B {\bf 423} (1994) 405.
  
\bibitem{Mohapatra:1993fe}
  R.~N.~Mohapatra and X.~Zhang,
  Phys.\ Rev.\ D {\bf 49} (1994) 1163
   Erratum: [Phys.\ Rev.\ D {\bf 49} (1994) 6246]
  [hep-ph/9307231].
  
\bibitem{Guth:1980zm}
  A.~H.~Guth,
  Phys.\ Rev.\ D {\bf 23} (1981) 347;
  D.~H.~Lyth and A.~Riotto,
  Phys.\ Rept.\  {\bf 314} (1999) 1
  [hep-ph/9807278].

\bibitem{Boucenna:2014uma}
  S.~M.~Boucenna, S.~Morisi, Q.~Shafi and J.~W.~F.~Valle,
  Phys.\ Rev.\ D {\bf 90} (2014) no.5,  055023
  [arXiv:1404.3198 [hep-ph]].
  
  \bibitem{Senoguz:2004ky}
  V.~N.~Senoguz and Q.~Shafi,
  Phys.\ Lett.\ B {\bf 596} (2004) 8
  [hep-ph/0403294].
  
\bibitem{Boubekeur:2005zm}
  T.~Asaka, K.~Hamaguchi, M.~Kawasaki and T.~Yanagida,
  Phys.\ Rev.\ D {\bf 61} (2000) 083512
  [hep-ph/9907559];
  L.~Boubekeur and D.~H.~Lyth,
  JCAP {\bf 0507} (2005) 010
    [hep-ph/0502047];
  K.~Kohri, C.~M.~Lin and D.~H.~Lyth,
  JCAP {\bf 0712} (2007) 004
  [arXiv:0707.3826 [hep-ph]];
  S.~Antusch and F.~Cefalˆ,
  JCAP {\bf 1310} (2013) 055
  [arXiv:1306.6825 [hep-ph]];
  S.~Antusch, D.~Nolde and S.~Orani,
  JCAP {\bf 1506} (2015) no.06,  009
  [arXiv:1503.06075 [hep-ph]];
  S.~Antusch and S.~Orani,
  JCAP {\bf 1603} (2016) no.03,  026
  [arXiv:1511.02336 [hep-ph]].
  
\bibitem{King:2014iia}
  S.~F.~King,
  JHEP {\bf 1408} (2014) 130
  [arXiv:1406.7005 [hep-ph]].


\bibitem{Antusch:2008gw}
  S.~Antusch, S.~F.~King, M.~Malinsky, L.~Velasco-Sevilla and I.~Zavala,
  Phys.\ Lett.\ B {\bf 666} (2008) 176
  [arXiv:0805.0325 [hep-ph]];
  S.~Antusch and D.~Nolde,
  JCAP {\bf 1310} (2013) 028
  [arXiv:1306.3501 [hep-ph]];
  S.~Antusch and D.~Nolde,
  JCAP {\bf 1509} (2015) no.09,  055
  [arXiv:1505.06910 [hep-ph]].

\bibitem{Antusch:2004hd}
  S.~Antusch, M.~Bastero-Gil, S.~F.~King and Q.~Shafi,
  Phys.\ Rev.\ D {\bf 71} (2005) 083519
  [hep-ph/0411298].


\bibitem{Murayama:1992ua}
  H.~Murayama, H.~Suzuki, T.~Yanagida and J.~Yokoyama,
  Phys.\ Rev.\ Lett.\  {\bf 70} (1993) 1912;
  J.~R.~Ellis, M.~Raidal and T.~Yanagida,
  Phys.\ Lett.\ B {\bf 581} (2004) 9
    [hep-ph/0303242];
  J.~R.~Ellis,
  Nucl.\ Phys.\ Proc.\ Suppl.\  {\bf 137} (2004) 190
  [hep-ph/0403247];
  F.~Bjorkeroth, S.~F.~King, K.~Schmitz and T.~T.~Yanagida,
  arXiv:1608.04911 [hep-ph];
  K.~Nakayama, F.~Takahashi and T.~T.~Yanagida,
  Phys.\ Lett.\ B {\bf 757} (2016) 32
    [arXiv:1601.00192 [hep-ph]];
  S.~Antusch, J.~P.~Baumann, V.~F.~Domcke and P.~M.~Kostka,
  JCAP {\bf 1010}, 006 (2010)
  [arXiv:1007.0708 [hep-ph]];
  S.~Khalil and A.~Sil,
  Phys.\ Rev.\ D {\bf 84} (2011) 103511
   [arXiv:1108.1973 [hep-ph]].

\bibitem{King:1997ia}
  S.~F.~King and Q.~Shafi,
  Phys.\ Lett.\ B {\bf 422} (1998) 135
  [hep-ph/9711288].

\bibitem{Zeldovich:1974uw}
  Y.~B.~Zeldovich, I.~Y.~Kobzarev and L.~B.~Okun,
  Zh.\ Eksp.\ Teor.\ Fiz.\  {\bf 67} (1974) 3
   [Sov.\ Phys.\ JETP {\bf 40} (1974) 1].

\bibitem{Kolb-Turner}
  E.~W.~Kolb and M.~S.~Turner,
  Front.\ Phys.\  {\bf 69} (1990) 1.

\bibitem{Ade:2015xua}
  P.~A.~R.~Ade {\it et al.} [Planck Collaboration],
  arXiv:1502.01589 [astro-ph.CO].



\end{thebibliography}
\end{document}